\documentclass{emulateapj}
\usepackage{lscape}

\slugcomment{ApJ, 2008, in press}

\begin{document}

\shortauthors{Gordon et al.}
\shorttitle{Aromatics in M101}

\title{The Behavior of the Aromatic Features in M101 HII Regions:
Evidence for Dust Processing}

\author{Karl~D.~Gordon\altaffilmark{1,2}, 
   Charles~W.~Engelbracht\altaffilmark{2}, 
   George~H.~Rieke\altaffilmark{2}, 
   K.~A.~Misselt\altaffilmark{2},
   J.-D.~T.~Smith\altaffilmark{2,3}, \&
   Robert~C.~Kennicutt, Jr.\altaffilmark{2,4}}

\altaffiltext{1}{Space Telescope Science Institute, Baltimore, MD 21218}
\altaffiltext{2}{Steward Observatory, University of Arizona, Tucson, AZ 85721}
\altaffiltext{3}{Ritter Astrophysical Research Center, University of
Toledo, Toledo, OH 43603}
\altaffiltext{4}{Institute of Astronomy, University of Cambridge,
Madingley Road, Cambridge CB3 0HA, UK} 

\begin{abstract} 
The aromatic features in M101 were studied spectroscopically and
photometrically using observations from all three instruments on the
Spitzer Space Telescope.  The global SED of M101 shows strong aromatic
feature (commonly called PAH feature) emission.  The spatially
resolved spectral and photometric measurements of the aromatic feature
emission show strong variations with significantly weaker emission at
larger radii.  We compare these variations with changes in the
ionization index (as measured by [NeIII]/[NeII] and [SIV/SIII], which
we probe over the ranges 0.03--20 and 0.044--15 respectively) and
metallicity (expressed as log(O/H) + 12, which ranges from 8.1 to
8.8). Over these ranges, the spectroscopic equivalent widths of the
aromatic features from 7 HII regions and the nucleus were found to
correlate better with ionization index than with metallicity.  This
implies that the weakening of the aromatic emission in massive star
forming regions is due primarily to processing of the dust grains in
these environments, not to differences in how they form (although
formation could still be important on a secondary basis).  The
behavior of the correlation between the aromatic feature equivalent
widths and ionization index can be described as a constant equivalent
width until a threshold in ionization index is reached ([NeIII]/[NeII]
$\sim 1$), above which the equivalent widths decrease with a power law
dependence.  This behavior for M101 HII regions is also seen for the
sample of starburst galaxies presented in the companion study of
\citet{Engelbracht08SB} which expands the range of
[NeIII]/[NeII] ratios to 0.03--25 and log(O/H)+12 values to 7.1--8.8.
The form of the correlation explains seemingly contradictory results
present in the literature.  The behavior of the ratios of different
aromatic features versus ionization index does not follow the
predictions of existing PAH models of the aromatic features implying a
more complex origin of the aromatic emission in massive star forming
regions.
\end{abstract}


\section{Introduction}
\label{sec_intro}

The interpretation of infrared measurements of star forming galaxies
rests on an understanding of the properties of the dust producing the
emission. In addition, the dependence of dust properties on
environment provides important clues to the dust composition and
formation.  An important constituent of the interstellar dust is
large, aromatic molecules/grains whose emission often dominates the
mid-infrared output of HII regions and galaxies \citep{Roche91,
Madden06, Smith07}.

The aromatic features were first discovered in observations of planetary
nebulae where a feature at 11.3~\micron\ was broader than the observed
atomic emission lines \citep{Gillett73}.  The number of these features
has grown to include many features with wavelengths between 3 and
18~\micron\ \citep{Werner04NGC7023, Smith04, Tielens05}.
The strongest aromatics are seen at 3.3, 6.2, 7.7, 8.6, 11.3, 12.7,
and 17.1~\micron\ and all the aromatics have been identified with C-H
and C-C bending and stretching modes of hydrocarbons containing
aromatic rings \citep{Tielens05}.  The shape and strength of the
aromatics have been seen to vary in single sources
\citep{Werner04NGC7023}, among various Galactic sources
\citep{vanDiedenhoven04}, and among galaxies \citep{Engelbracht05SB,
Madden06, Wu06}.  

A number of different materials have been proposed as the carriers of
these features, including Hydrogenated Amorphous Carbon
\citep[HAC,][]{Duley83}, Quenched Carbonaceous Composites
\citep[QCC,][]{Sakata84}, Polycyclic Aromatic Hydrocarbons
\citep[PAHs,][]{Allamandola85}, coal \citep{Papoular89}, and
nanodiamonds \citep{Jones00}.  The leading candidate material for the
aromatics is PAH molecules and they are often included in the modified
``astronomical PAHs'' form in dust grain models \citep{Li01, Zubko04}.
We refer to these features as the aromatic features to allow for
discussions of the observations without making an assumption about the
material that is responsible for them.  We refer the interested
reader to \citet{Peeters04} for a detailed discussion of these
features.

The observed strengths of the aromatic features do not vary much in
most normal luminosity galaxies with approximately solar metallicities
\citep{Roche91, Lu03, Smith07} or massive starbursts
\citep{Brandl06SB}.  The strength of these features
do weaken significantly in active galactic nuclei \citep{Roche91,
Lutz98, Smith07} and star bursting galaxies with fairly low
metallicities
\citep{Roche91, Engelbracht05SB, Galliano05, Madden06, Wu06,
Engelbracht08SB}.  The variations seen in active galactic nuclei are
not considered for this paper, which only discusses those clearly
associated with nearby massive star formation (e.g., starburst
galaxies and HII regions).  Observations with the Infrared Space
Observatory (ISO) found that the strength of the aromatic features was
correlated with a galaxy's metallicity (probed by log(O/H)+12) and
ionization (probed by [NeIII]/[NeII]), but were unable to determine
which was the dominant effect \citep{Madden06}.  The correlations are
such that the aromatic features are weakest for low metallicities and
high ionizations.  

If the correlation is better with metallicity than ionization, then
this implies that the formation of the aromatics (from the constituent
atoms in circumstellar or interstellar environments) happens less
efficiently than that of small grains as the metallicity decreases.
On the other hand, if the better correlation is with ionization, then
this implies that there is increased processing of the aromatic
feature carriers (modification and/or destruction in the immediate
environment) as the radiation field becomes harder.  Throughout this
paper, we use the metallicity as a probe of the formation of aromatics
from the constitute atoms and ionization as a probe of the processing
of the aromatics in the immediate environment.  The metallicity and
ionization in massive star forming regions is roughly correlated
making it nontrivial to determine which is the dominate correlation.

Observations with Spitzer have refined the observational picture of
the variations of the aromatic features among galaxies, but have yet
to provide a clear picture of the origin of the variations.  Evidence
that the aromatic features are absent at the lowest 
metallicities probed by starburst galaxies was provided by the
IRS (Infrared Spectrograph) spectrum of SBS~0335-052 which has a
log(O/H)+12=7.2 
\citep{Houck04SBS}.  The weakness/absence of aromatic features at
metallicities below log(O/H)+12=8.2 in a sample of starburst galaxies was shown by
\citet{Engelbracht05SB} using IRAC (Infrared Camera) and MIPS (Multiband
Imaging Photometry for Spitzer) photometry confirmed in a
few cases with IRS spectroscopy.  \citet{Wu06} studied a small sample
of Blue Compact Dwarf galaxies (including SBS~0335-052) with IRS
spectroscopy and found similar correlations between
aromatic feature equivalent width and metallicity or radiation field
hardness.  \citet{Brandl06SB} found that the equivalent width of the
aromatic features did not vary significantly over a factor of 10 in
[NeIII]/[NeII] for a sample of starburst galaxies.  Using the HII
classified galaxies in the SINGS (Spitzer Nearby Galaxies Survey)
sample \citet{Smith07} found 
variations in the summed strength of the aromatic features when
ratioed to the total IR flux with some indication of a threshold at
log(O/H)+12=8.1.  The metallicity threshold in the aromatic feature
strengths was confirmed by \citet{Draine07} where the broadband
spectra energy distributions (SEDs)
of the SINGS galaxies were modeled.  These models indicated a fairly
abrupt reduction by a factor of three occurs in the fraction of PAHs
near a metallicity of log(O/H)+12 = 8.1.

All this variation among galaxies raises the question: do the
aromatics vary significantly among HII regions in a single galaxy?
The study of the simpler environments of HII regions may also present
a clearer picture of the cause of the aromatic feature variations as
HII regions have much more homogeneous physical conditions than whole
galaxies.  The ideal galaxy to answer this question is the giant
spiral galaxy M101.  M101 is a large ($> 30$\arcmin diameter), face-on
spiral galaxy (SABcd) at a distance of 6.7~Mpc \citep{Freedman01Ho}.  It has
one of the largest metallicity gradients known with metallicities
[log(O/H) + 12] ranging from 8.8 in the nucleus to 7.4 in an HII
region 41~kpc from the center \citep{Zaritsky94, vanZee98,
Kennicutt03, Bresolin06}.  In addition, there
are high quality electron temperature and metallicity determinations
for a sample of M101 HII regions with metallicities from 7.4 to 8.8
\citep{Kennicutt03, Bresolin06}.  This large apparent size, the
numerous HII regions, the large metallicity gradient, and the high
quality metallicity measurements make M101 ideal for studying the
dependence of 
the aromatics on the local conditions, specifically metallicity and
ionization.

\section{Data}
\label{sec_data}

The high sensitivity and spatial resolution provided by all three
instruments on Spitzer allow individual HII regions in M101 to be
studied in detail.  Targeted spectroscopic observations of a handful
of HII regions and the nucleus were taken with the Infrared
Spectrograph \citep[IRS,][]{Houck04IRS}.  Imaging of M101 was obtained
with the Infrared Array Camera \citep[IRAC,][]{Fazio04IRAC} at 3.6,
4.5, 5.8, 8.0~\micron\ and the Multiband Imaging Photometry for
Spitzer \citep[MIPS,][]{Rieke04MIPS} at 24, 70, and 160~\micron.  The
IRAC and MIPS images allow the general behavior of the aromatics to be
studied over the entire galaxy and the IRS spectra allow for the
behavior of the aromatics to be probed in detail for selected HII
regions.  Past studies of M101 in the infrared have concentrated on
radial gradients as the spatial resolution was not high enough to
study any but the few brightest HII regions \citep[eg.][]{Hippelein96,
Popescu05M101}.

\begin{figure*}[tbp]
\plotone{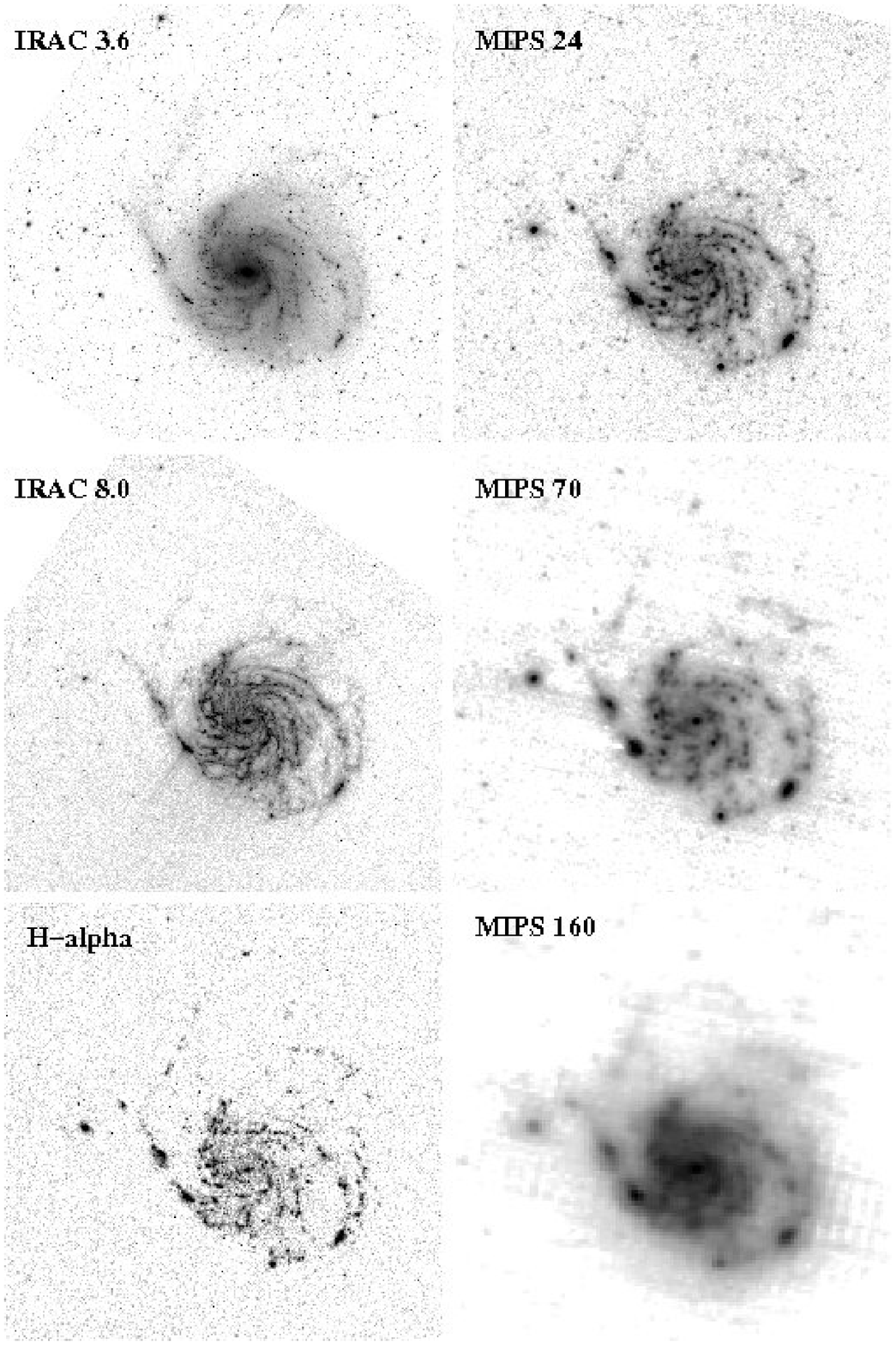}
\caption{IRAC, MIPS, and H$\alpha$ images of M101 at full resolution.
The IRAC 4.5 and 5.8~\micron\ images have not been shown as they look
very similar to the IRAC 3.6 and 8.0~\micron\ images, respectively.
The field-of-view of the images is $30\arcmin \times 30\arcmin$.  The
H$\alpha$ image \citep{vanZee98} is included to show the locations of
HII regions.  The streaking seen SE of M101 at 8~\micron\ and roughly
East-West at 70 and 160~\micron\ are residual instrumental artifacts.
\label{fig_m101_fullres} }
\end{figure*}

Preliminary results from the IRAC and MIPS images were presented by
\citet{Gordon05NewViews} and from the IRS spectroscopy in
\citet{Gordon06IRDiag}.  This work focuses on the intense star
forming HII regions in a single galaxy and is complementary to the
Spitzer program on starburst galaxies led by Charles Engelbracht
\citep{Engelbracht05SB, Engelbracht05NewViews, Engelbracht08SB} 
which focuses on the properties whole galaxies undergoing intensive
star formation.

\subsection{IRAC and MIPS Images}

The IRAC images of M101 were obtained on 8 March 2004.  Two maps were
taken separated by a few hours to allow for asteroid 
rejection.  Each map consisted of $13 \times 13$ tiles with 150\arcsec
offsets (1/2 the array width) between tile positions in both array
dimensions .  Each exposure was taken in high dynamic range (HDR) mode
resulting in two 
images, one with a 10.4 sec exposure time and one with a 0.4 second
exposure time.  The images were reduced with the Spitzer Science
Center (SSC) pipeline S13.2.0
and combined using MOPEX program (v030106).  As part of the mosaicking,
overlapping regions of adjacent images were used to correct for bias
drifts in each image.  The calibration of the images was corrected to
an infinite aperture (true surface brightness units) using the values
given by \citet{Reach05}.  The final mosaics have exposure times of
$\sim$85 seconds/pixel and are shown in Figure~\ref{fig_m101_fullres}.

The MIPS images of M101 were obtained on 10-11 May 2004.  Like the
IRAC images, two maps were taken separated by a few hours for asteroid
rejection.  Each map consisted of 14 medium rate 0.75 degree scan legs
with a cross scan offset of 148\arcsec.  We used version 3.06 of the
MIPS Data Analysis Tool (DAT)
\citep{Gordon05DAT} to do the basic processing and final mosaicking of
the individual images.  Extra processing steps before mosaicking used
programs written specifically to improve the reduction of large,
well-resolved galaxies.  At 24~\micron\ the extra steps included
readout offset correction, scan mirror dependent flat fields, array
averaged background subtraction (using a low order polynomial fit to
the data in each scan leg but excluding M101), and rejection of the
1st 5 images in each scan leg due to bias boost transients.  At 70 and
160~\micron\ the extra steps were a pixel dependent background
subtraction for each map (using a low order polynomial fit to the data
in each scan leg, again excluding M101) and a spatial cosmic ray
cleaning (160~\micron\ only).  The calibration of the MIPS bands was
done using the latest, official factors \citep{Engelbracht07MIPS24,
Gordon07MIPS70, Stansberry07MIPS160}.  The 5$\sigma$ depth achieved
in a single beam by these
observations is 0.050, 0.53, and 0.74 MJy sr$^{-1}$ for
beam FWHM of 6$\arcsec$, 18$\arcsec$, and 40$\arcsec$ and exposure
times of $\sim$200, $\sim$80, and $\sim$18 seconds/pixel for 24, 70,
and 160~\micron, respectively.  The final mosaics are shown in
Figure~\ref{fig_m101_fullres}.  The streaking seen along the scan
direction at 70 \& 160~\micron\ is caused by residual instrumental
artifacts and only produces a significant worsening of the noise at
70~\micron.  The online MIPS SENS-PET program predicts approximately the
observed noise at 24 \& 160~\micron\, but a factor of 3 lower noise
than observed for the 70~\micron\ observations.  This difference is
likely due to the residual streaking seen along the scan mirror
direction which can be suppressed by combining scan maps
taken at different scan angles \citep{Tabatabaei07}.  Additional MIPS
scan maps of M101 are planned and will be the subject of a future paper.

\begin{figure*}[tbp]
\epsscale{1.15}
\plotone{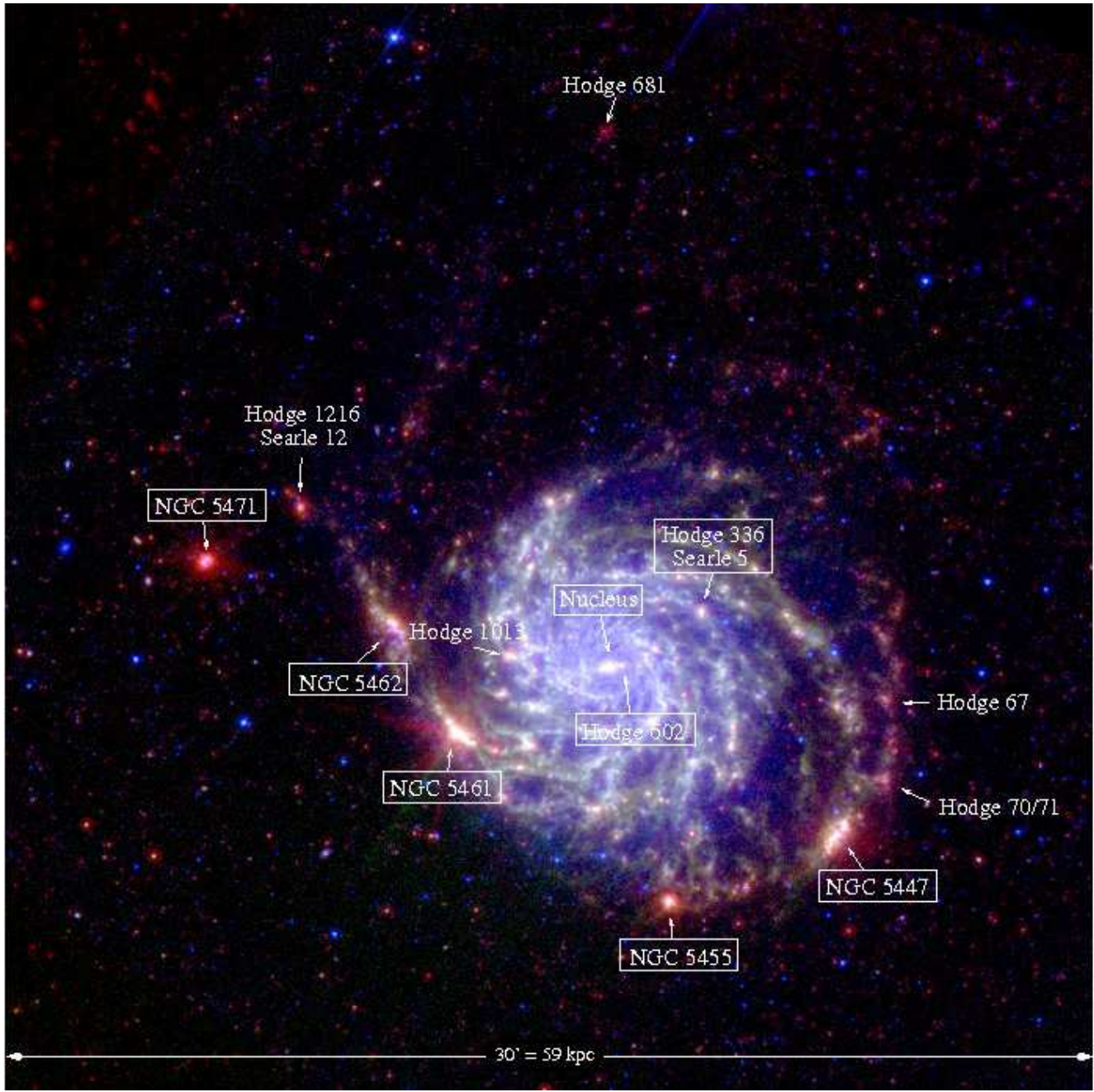}
\caption{A 3 color image of M101 is shown annotated with the locations
of HII regions with measured metallicities.  The red, green, and blue
channels are filled with the MIPS 24~\micron, IRAC 8~\micron, and IRAC
3.6~\micron\ images, respectively.  The regions that were
observed with IRS are identified as having a box around their names.
\label{fig_m101_truecol} }
\end{figure*}

\subsection{IRS Spectra}

\begin{figure*}[tbp]
\epsscale{1.15}
\plotone{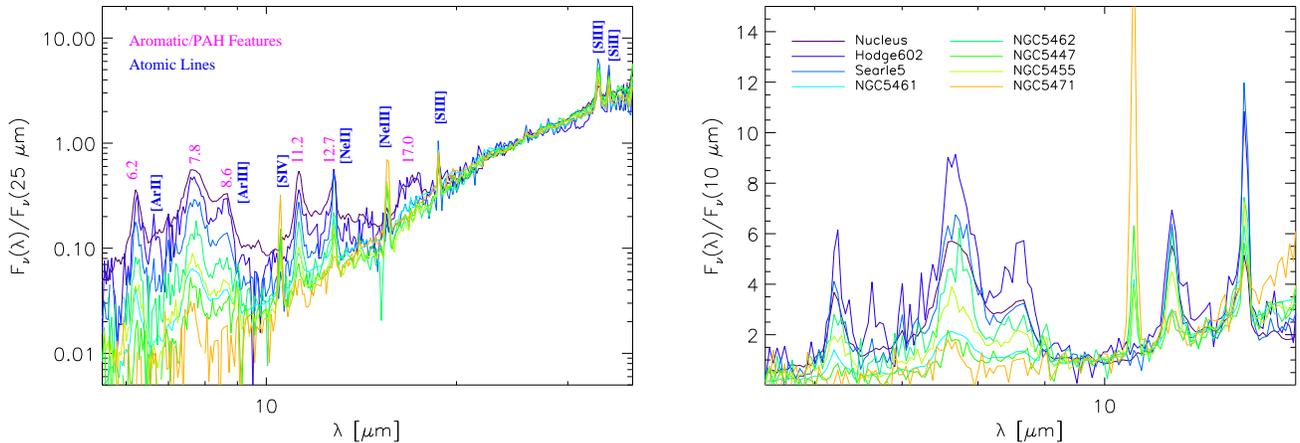}
\caption{The IRS spectrum for all 8 regions with targeted IRS
spectroscopy are shown.  The left plot show the full spectra
normalized to their average flux in the 20--30~\micron\ region and on a log
scale.  The
locations of aromatic and atomic emission lines are given.  The right
plot shows the 5--15~\micron\ portion of the spectra normalized to
their average flux at 10~\micron\ and on a linear scale.  The legend
identifying the regions is given in this plot.
\label{fig_m101_spec} }
\end{figure*}

\begin{deluxetable*}{llcccc}
\tablewidth{0pt}
\tabletypesize{\small}
\tablecaption{M101 Regions with measured metallicities \label{tab_m101_regions}}
\tablehead{\colhead{Name} & \colhead{Other Names} & \colhead{Coordinates} &
   \colhead{R/R$_o$\tablenotemark{a}} & \colhead{12+log(O/H)} & \colhead{IRS Spectra} }
\startdata
Nucleus   &           & 14 03 12.48 54 20 55.4 & 0.00 & $8.76 \pm 0.06$ & X \\
Hodge~602  &           & 14 03 10.22 54 20 57.8 & 0.02 & $8.76 \pm 0.06$ & X \\
Hodge~1013 &           & 14 03 31.39 54 21 14.5 & 0.19 & $8.71 \pm 0.05$ & \\
Searle~5   & Hodge~336  & 14 02 55.05 54 22 26.6 & 0.21 & $8.55 \pm 0.16$ & X \\
NGC~5461   & Hodge~1105 & 14 03 41.36 54 19 04.9 & 0.32 & $8.50 \pm 0.03$ & X \\
NGC~5447   & Hodge~128  & 14 02 28.18 54 16 26.3 & 0.55 & $8.45 \pm 0.04$ & X \\
NGC~5462   & Hodge~1170 & 14 03 53.19 54 22 06.3 & 0.44 & $8.37 \pm 0.07$ & X \\
NGC~5455   & Hodge~409  & 14 03 01.13 54 14 28.7 & 0.46 & $8.25 \pm 0.03$ & X \\
Hodge~67   &           & 14 02 19.92 54 19 56.4 & 0.54 & $8.14 \pm 0.08$ & \\
Hodge~70/71 &          & 14 02 20.50 54 17 46.0 & 0.57 & $8.10 \pm 0.12$ & \\
Searle~12  & Hodge~1216 & 14 04 11.11 54 25 17.8 & 0.67 & $8.16 \pm 0.03$ & \\
NGC~5471   & NGC~5471-D & 14 04 29.35 54 23 46.4 & 0.80 & $8.09 \pm 0.03$ & X \\
Hodge~681  &           & 14 03 13.64 54 35 43.0 & 1.03 & $7.92 \pm 0.09$ & \\
\enddata
\tablenotetext{a}{R$_o = 14.42\arcmin$}
\end{deluxetable*}

Low resolution IRS spectra from 5-38~\micron\ of 7 HII regions and the
nucleus were taken in the spectral mapping mode with full slit width
steps perpendicular to the slit.  The locations of these 7 regions as
well as other HII regions with measured electron temperatures and
metallicities from 
\citet{Kennicutt03} are shown in Fig.~\ref{fig_m101_truecol} and their
basic details are given in Table~\ref{tab_m101_regions}.  The ShortLow
spectra (5--15~\micron) were taken with 6 steps and the LongLow
(15--38~\micron) were taken with 2 
steps so that both spectral apertures covered the same regions.  The
metallicities [12+log(O/H)] are taken from table~5 of
\citet{Kennicutt03} except for the values for the nucleus and
Hodge~602, which are the constant term in their equation~5.  The
ShortLow and LongLow observations were taken approximately 3 months
apart so the orientations of the two slits were approximately the same
on the sky.

\begin{figure*}[tbp]
\epsscale{1.15}
\plotone{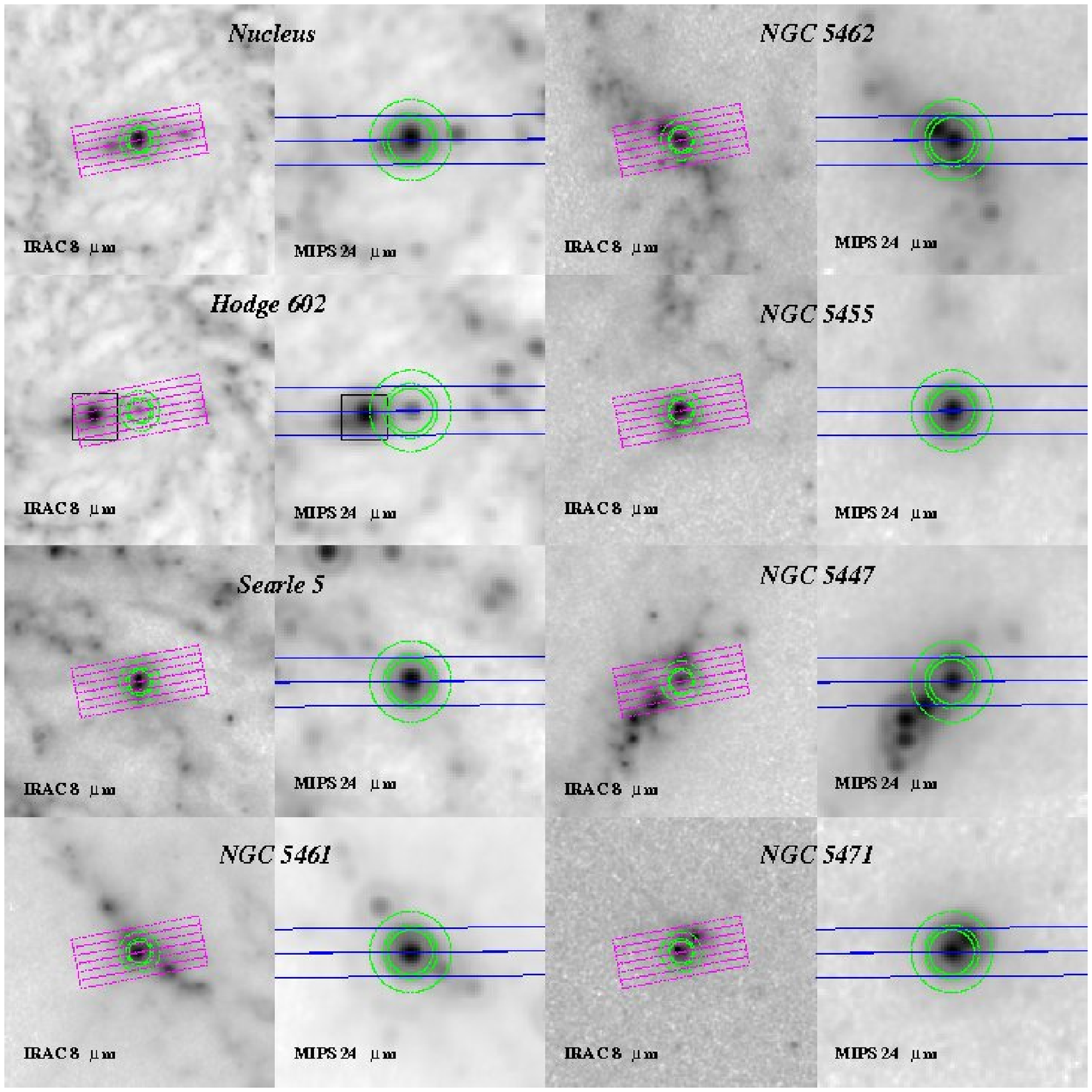}
\caption{Cutouts centered on each of the 8 regions with targeted IRS
spectroscopy are shown.  Each cutout is $2\arcmin \times 2\arcmin$ and
shows the circular
extraction aperture (and sky 
annulus) at that wavelength and the outline of the ShortLow and
LongLow slits on the IRAC 8~\micron\ and MIPS 24~\micron\ images,
respectively.  For Hodge~602, the nearby nuclear region was masked
(black square shown) to allow for an accurate sky measurement.
\label{fig_m101_ext_regions} }
\end{figure*}

These spectral mapping observations were reduced with the SSC pipeline
(version S13/S14) and combined using CUBISM program \citep{Smith07cubism}.
Spectra were extracted from the cubes for each HII region and 
the nucleus using an aperture with a radius of 10\arcsec\ and sky
annuli from 12 to 18\arcsec\ referenced at 24~\micron.  The size of
the aperture and sky annuli were varied linearly with wavelength to
account roughly for the changing diffraction limited
point-spread-function (PSF), with a
minimum object aperture radius of 5\arcsec\ to account of the finite
size of the HII regions.  The outlines of the individual IRS slits and
extraction apertures are shown in Fig.~\ref{fig_m101_ext_regions}
superimposed on cutouts of the IRAC~8~\micron\ and MIPS~24~\micron\
images centered on each HII region.  There are issues of crowding with
nearby HII regions for 
some of the targets, but for the most part the targets are fairly well
separated.  The measured noise in the sky annuli region was used to
compute the 
uncertainty on each point in the spectrum.

The six spectral segments for each target were merged into a single
spectrum using the overlap regions to adjust the level of each segment
to be consistent.  The overall level of the spectra was set by
requiring each spectrum to be consistent with the MIPS 24~\micron\
measurement given in Table~\ref{tab_m101_regphot}.  The main
adjustment is at the transition between the ShortLow and LongLow
spectra.  This is not surprising given the observing strategy where
the region is sampled with 6 ShortLow positions, but only 2 LongLow
positions.  The resulting merged spectra are shown in
Fig.~\ref{fig_m101_spec}.

\subsection{Resolution Matching}
\label{sec_res_match}

One of the challenges of working with Spitzer data is the large change
in PSF with wavelength as Spitzer is diffraction limited at
wavelengths longer than $\sim$6~\micron.  The FWHM (full width half
maximum) of the Spitzer PSF 
varies from around $2\arcsec$ for the IRAC bands to $5\arcsec$,
$18\arcsec$, and $40\arcsec$ for the MIPS 24, 70, and 160~\micron\
bands.  To allow for accurate comparisons between wavelengths, it is
important to convolve all images of interest to the lowest common
resolution.  Without this step, it is not easy to compare extended
sources at different Spitzer bands given that the physical region
probed changes with wavelength.  The Spitzer PSFs are much more
complex than simple Guassians with pronounced Airy rings and
diffraction spikes.  This makes it definitely
non-optimal to use simple Gaussian kernels to achieve the desired
common resolution.  It is also
possible to put a set of images on a common resolution by convolving
each image with all the other image PSFs, but this clearly creates a
lower resolution common PSF than the method detailed below.

The optimal method in the case of complex PSFs is to construct
convolution kernels that transform the input PSF to the desired
common resolution PSF.  Theoretically, the desired convolution kernel
can be constructed using 2D Fast Fourier Transforms (FFTs) and is just
\begin{equation}
K(x,y) = \mathrm{FFT}^{-1} \left\{ \frac{\mathrm{FFT} 
   \left[ \mathrm{PSF}_2(x,y) \right]}
   {\mathrm{FFT} \left[ \mathrm{PSF}_1(x,y) \right]} \right\}
\end{equation}
where $\mathrm{PSF}_1$ is the input image PSF, $\mathrm{PSF}_2$ is the
desired common resolution PSF, and K is the convolution kernel that
transforms $\mathrm{PSF}_1$ to $\mathrm{PSF}_2$.  Unfortunately, the
high frequency numeric noise in $\mathrm{PSF}_1$ is greatly amplified
and the resulting kernel is dominated by noise.  The solution is to
add a clamping function that attenuates the high frequency noise
$\mathrm{FFT}\left[ \mathrm{PSF}_1(x,y) \right]$.  As the dominant
signal we are interested in is radial, we adopt a radial Hanning truncation
function, $W_h(\omega)$, with the form
\begin{equation}
W_h(\omega) = \left\{ \begin{array}{ll}
    \frac{1}{2} \left[ 1 + \cos \left( \frac{2 \pi \omega}{\omega_o}
      \right) \right] & \omega \le \omega_o, \\
    0 & \omega > \omega_o \\ \end{array} \right.
\label{eq_wh}
\end{equation}
where $\omega$ is radial spatial frequency and $\omega_o$ is the
cutoff radial spatial frequency \citep{Brigham88}.  The 
revised equation for the desired convolution kernel is then
\begin{equation}
K(x,y) = \mathrm{FFT}^{-1} \left\{ \frac{\mathrm{FFT} 
   \left[ \mathrm{PSF}_2(x,y) \right]}
   {W_h(\omega) \mathrm{FFT} \left[ \mathrm{PSF}_1(x,y) \right]} \right\}.
\end{equation}
The value of $w_o$ in eq.~\ref{eq_wh} is chosen to be as large as
possible while still suppressing the high frequency noise in
$\mathrm{FFT} \left[ \mathrm{PSF}_1(x,y) \right]$.

\begin{figure}[tbp]
\epsscale{1.15}
\plotone{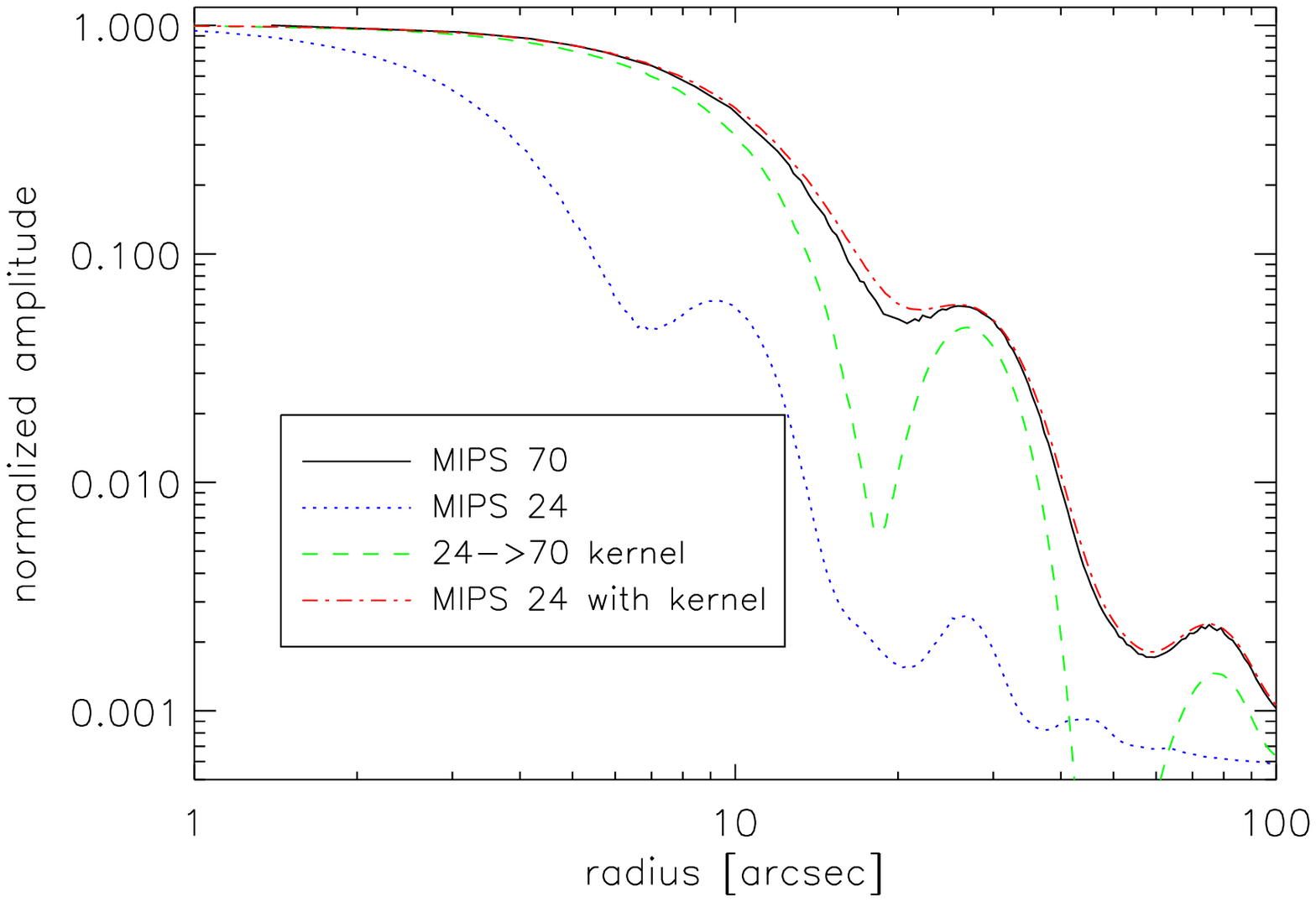}
\caption{The radial
profiles of the 24~\micron\ PSF, the 24-to-70~\micron\ 
kernel, the 70~\micron\ PSF, and the convolution of the 24~\micron\
PSF and the 24-to-70~\micron\ kernel are shown.
\label{fig_conv_example} }
\end{figure}

This method of creating convolution kernels creates a set of images
with a common PSF of the lowest resolution in the set.  For this
method to work, the PSFs of each image have to be known to a
reasonable degree and convolution kernels created for each unique PSF
in the image set.  The value of $\omega_o$ is set as large as possible for
each kernel by visual inspection of the convolution of
$\mathrm{PSF}_1(x,y)$ with $K(x,y)$.  The goal is for this
convolution to reproduce $\mathrm{PSF}_2(x,y)$ with the highest
possible accuracy while minimizing the noise injected into the operation.  An
example showing the performance of the 24-to-70~\micron\ convolution
kernel is given in Fig.~\ref{fig_conv_example}.  The close (but not
perfect) correspondence between the 70~\micron\ PSF and the
convolution of the 24~\micron\ PSF with the 24-to-70~\micron\ kernel
is a validation of the accuracy of this method.

For the work on M101, we created convolution kernels that transform
the IRAC PSFs to the MIPS 24~\micron\ PSF for use in
\S\ref{sec_photo_hII} and kernels that transform the IRAC, MIPS
24~\micron\, and MIPS 70~\micron\ PSFs to the MIPS 160~\micron\ PSF for
use in \S\ref{sec_morphology}.  Ideally, we would have done the same
to match the resolutions of each IRS wavelength to the
resolution of the longest IRS wavelength.  Unfortunately, the IRS PSFs
are not nearly as well characterized as the IRAC and MIPS PSFs and our
small regions mapped with IRS would introduce significant edge effects
in any convolutions.  Instead, we varied the extraction
apertures to account for the changing IRS PSF with wavelength.

\section{Results}

With the data presented in this paper, the behavior of the aromatic
features in M101 can be probed both photometrically and
spectroscopically.  The IRAC and MIPS total M101 fluxes are combined
with previous measurements to probe the global SED of M101
(\S\ref{sec_global_sed}).  Using the IRS spectra of the targeted HII
regions, we probe the aromatic feature behavior in detail in
\S\ref{sec_spec_aromatic}.  Returning to the imaging data, the
behavior of the aromatics for an expanded sample of regions is
studied (\S\ref{sec_photo_hII}) and, finally, the overall infrared
morphology is used to gain further insight into the aromatics in M101
(\S\ref{sec_morphology}).  Combining both the spectroscopic and
photometric approaches produces a richer picture of the aromatic
feature behavior than is possible with either approach separately.

\subsection{Global SED}
\label{sec_global_sed}

The global IR SED of M101 is useful to put this galaxy in context with
other galaxies and to probe the overall aromatic and dust content.
The IRAC and MIPS global fluxes were
measured using a circular aperture with a radius of $13\farcm 3$ and a
sky annulus from $14\farcm 2$ to $16\farcm 7$.  The uncertainties on
these fluxes 
are dominated by the uncertainties in the absolute calibrations.  The
IRAC absolute calibration uncertainties are taken as 5\% which is
larger than the point source uncertainty of $\sim$2\% \citep{Reach05}
as the IRAC scattered light behavior makes the extended source
calibration less accurate.  The MIPS
absolute uncertainties are 4\% \citep{Engelbracht07MIPS24}, 5\%
\citep{Gordon07MIPS70}, and 12\% \citep{Stansberry07MIPS160}.  
We have tabulated previous IR measurements of the total M101 emission
in Table~\ref{tab_global_sed}.  We have included the ISOPHOT
\citep{Lemke96} flux from 
\citet{Stickel04} but not the fluxes given by \citet{Tuffs03} as these
fluxes are known to be lower limits.  The Midcourse Space Experiment
\citep[MSX,][]{Cohen01MSX} observed M101 in four mid-infrared bands
\citep{Kraemer02}, but only had a high signal-to-noise detection in
MSX Band A.  The MSX Band A flux was measured using the same apertures
as the IRAC and MIPS fluxes from the image available from NED.  The
uncertainty on this measurement was set to 20\% to account for the
difficulty in sky subtraction for these data.

\begin{deluxetable*}{lccccc}
\tablewidth{0pt}
\tablecaption{M101 Global Fluxes \label{tab_global_sed}}
\tablehead{ & \colhead{Wavelength} & \colhead{Bandwidth} & \colhead{Flux} & & \\
 \colhead{Band} & \colhead{[\micron]} & \colhead{[\micron]} & \colhead{[Jy]} & 
 \colhead{Origin} & \colhead{Reference} }
\startdata
J       & 1.235 & 0.162 & $3.94 \pm 0.12$  & 2MASS & 1 \\
H       & 1.662 & 0.251 & $4.88 \pm 0.19$  & 2MASS & 1 \\
K       & 2.159 & 0.262 & $4.16 \pm 0.20$  & 2MASS & 1 \\
IRAC1   & 3.550 & 0.681 & $2.84 \pm 0.14$  & Spitzer/IRAC & 2 \\
IRAC2   & 4.493 & 0.872 & $1.76 \pm 0.09$  & Spitzer/IRAC & 2 \\
IRAC3   & 5.731 & 1.250 & $3.69 \pm 0.18$  & Spitzer/IRAC & 2 \\
LW2     & 6.75  & 3.5   & $6.03 \pm 0.442$ & ISO/ISOCAM & 3 \\
IRAC4   & 7.872 & 2.526 & $7.26 \pm 0.36$  & Spitzer/IRAC & 2 \\
MSXA    & 8.276 & 3.362 & $8.86 \pm 1.77$  & MSX & 2 \\
IRAS12  & 12    & 7.0   & $6.2 \pm 0.93$   & IRAS & 4 \\
COBE12  & 12    & 6.48  & $< 16.9$         & COBE/DIRBE & 5 \\
LW3     & 15.0  & 6.0   & $5.42 \pm 0.501$ & ISO/ISOCAM & 3 \\
MIPS24  & 23.7  & 4.7   & $10.5 \pm 0.4$  & Spitzer/MIPS & 2 \\
IRAS25  & 25    & 11.2  & $11.8 \pm 1.77$  & IRAS & 4 \\
COBE25  & 25    & 8.60  & $12.2 \pm 2.44$  & COBE/DIRBE & 5 \\
IRAS60  & 60    & 32.5  & $88.0 \pm 13.2$  & IRAS & 4 \\
COBE60  & 60    & 27.84 & $82.8 \pm 16.6$  & COBE/DIRBE & 5 \\
MIPS70  & 71.0  & 19.0  & $116.3 \pm 7.0$ & Spitzer/MIPS & 2 \\
IRAS100 & 100   & 31.5  & $252.8 \pm 37.9$ & IRAS & 4 \\
COBE100 & 100   & 32.47 & $259.0 \pm 51.8$ & COBE/DIRBE & 5 \\
COBE140 & 140   & 39.53 & $< 530.0$        & COBE/DIRBE & 5 \\
MIPS160 & 156   & 35.0  & $405.3 \pm 35.0$ & Spitzer/MIPS & 2 \\
ISO170  & 170   & 90.2  & $495.8 \pm 74.4$ & ISO/ISOPHOT & 6 \\
COBE240 & 240   & 95.04 & $< 497.5$        & COBE/DIRBE & 5 \\
\enddata
\tablerefs{(1) \citet{Jarrett03}; (2) this work; (3) \citet{Roussel01};
 (4) \citet{Rice88} (5) \citet{Odenwald98} (6) \citet{Stickel04}}
\end{deluxetable*}

\begin{figure*}[tbp]
\epsscale{0.85}
\plotone{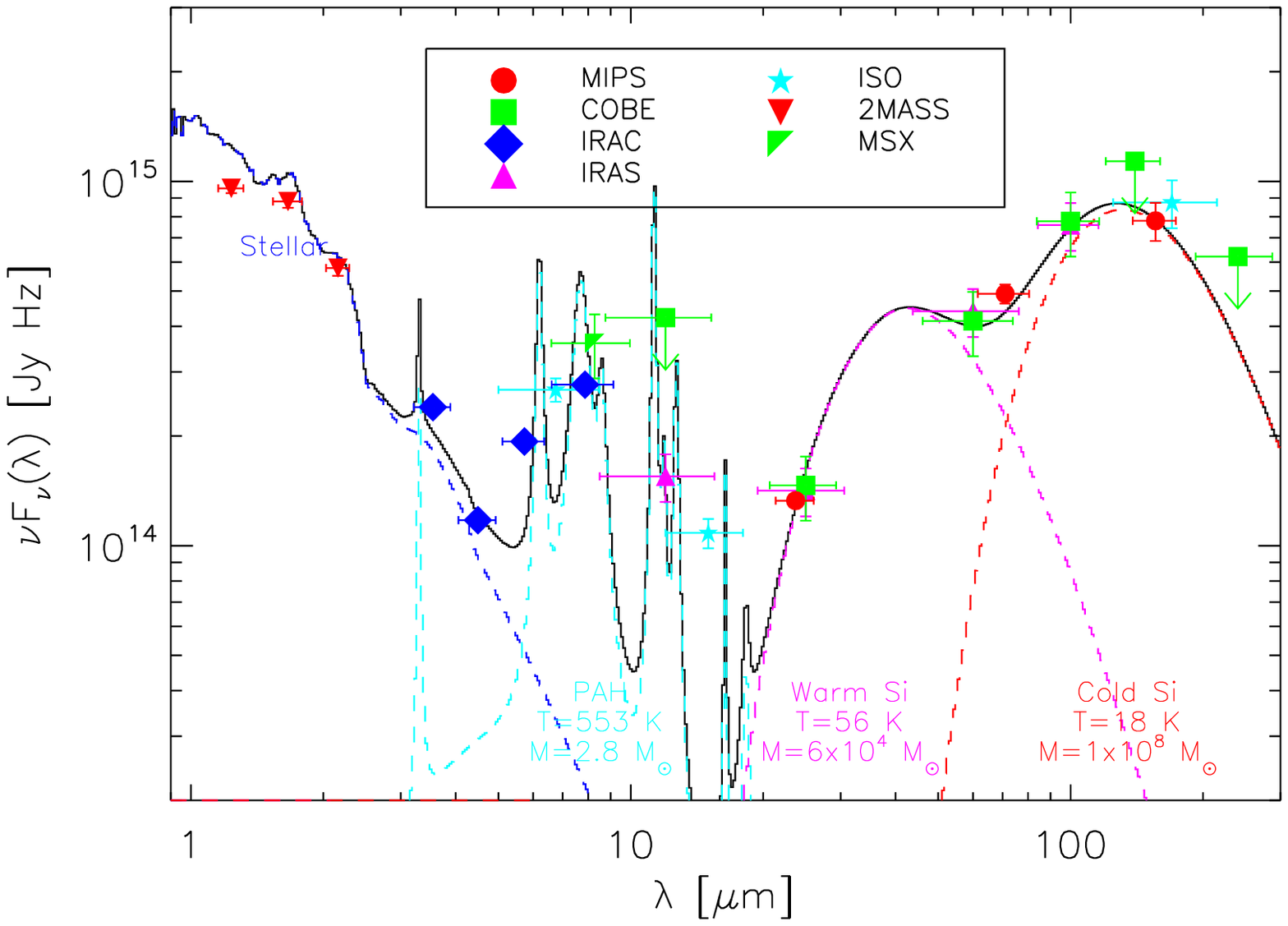}
\caption{The IR SED of M101 is shown from the new IRAC and MIPS
observations presented here as well as measurements taken from the
literature.  The COBE/DIRBE upper limits are 3$\sigma$.  The horizontal
error bars give the widths of the bands.  A simple stellar population
plus 3 component dust emission model fit is shown.  Each dust
component is modeled with a single grain size and the temperature
given is the equilibrium temperature.  The PAH component is
illustrative only as it is assumed to be in thermal equilibrium which
is an approximation as this component is known to be heated
stochastically.
\label{fig_global_sed} }
\end{figure*}

The global IR SED of M101 is shown in Fig.~\ref{fig_global_sed} and
looks similar to many of the Spitzer Nearby Galaxies Survey
\citet[][SINGS,]{Kennicutt03SINGS} galaxy global SEDs \citep{Dale05,
Dale07, Draine07}.  We fit the global SED of M101 with the combination
of a simple stellar population and a three component dust grain
emission model.  The details of this modeling can be found in
\citet{Marleau06}.  While this model is not unique, it is illustrative
of the dust properties implied by the SED.  The global IR SED of M101
includes strong aromatic emission features as well as emission from
warm (T~$\sim 56$~K) and cold (T~$\sim 18$~K) dust.  The total dust
mass of $1\times 10^{8}$~M$_{\sun}$ is dominated by the coldest
component and is much larger than the previous estimate of $4.8\times
10^{6}$~M$_{\sun}$ based on IRAS data alone \citep{Young89}.  Using
our new measurement of the dust mass, the M101 gas-to-dust ratio is
$\sim$300 using a HI mass of $2.5\times 10^{10}$~M$_{\sun}$
\citep{Bosma81} and a H$_2$ mass of $5.1\times 10^{9}$~M$_{\sun}$.
This new gas-to-dust ratio is much smaller than the previous
measurement of 2440 by \citet{Devereux90}.  This is in accordance with
their suspicion that there was an order of magnitude more dust which
was too cold for 
IRAS to measure.  The dust-to-gas ratio we measure is consistent with
that expected for a galaxy with a global metallicity of 8.4
\citep{Kennicutt03, Moustakas06} from the work of \citet{Draine07} on
SINGS global SEDs.

The global obscured SFR can be measured using the total IR luminosity
and the calibration given by \citet{Kennicutt03}.  The total IR luminosity of
M101 determined using the TIR (total IR) formulation \citep{Dale02} is $2.5
\times 10^{43}$ ergs s$^{-1}$ which gives a SFR of 1.1
$M_{\sun}$~year$^{-1}$.  The global obscured SFR can also be measured
using the MIPS 24~\micron\ luminosity ($\lambda F(24~\micron) 4\pi
 d^2$, where d = distance) which is $7.05 \times 10^{43}$ ergs
s$^{-1}$.  This  gives a SFR 
of 1.8 and 1.1 $M_{\sun}$~year$^{-1}$ using the calibrations of
\citet{Alonso06} or \citet{Calzetti07}, respectively.

\subsection{Spectroscopic Measures of Aromatics}
\label{sec_spec_aromatic}

\begin{figure*}[tbp]
\epsscale{1.1}
\plotone{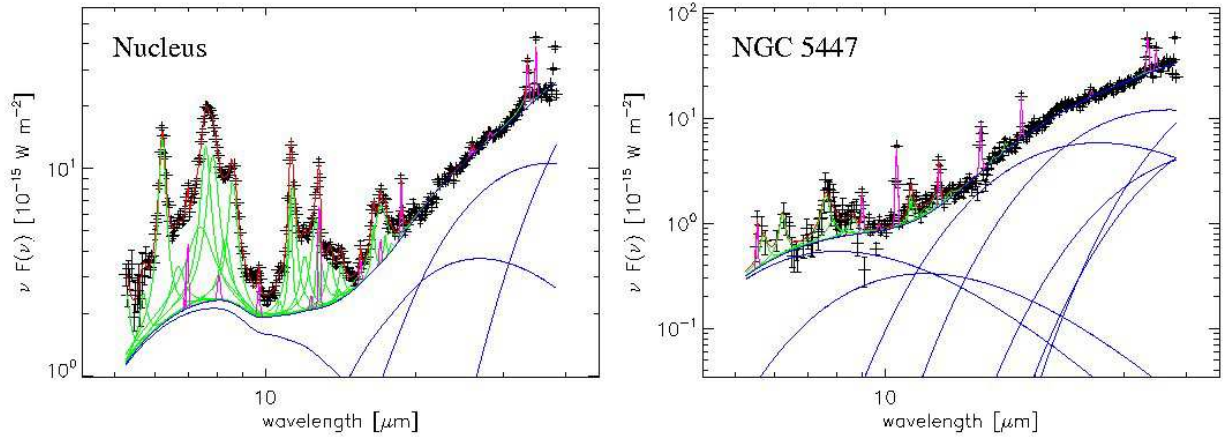}
\caption{The IRS spectra for the Nucleus and NGC~5447 are shown along
with the PAHFIT results.  The total PAHFIT spectrum is shown as well
as the decomposition of the total (red) into dust continuum (blue),
atomic emission lines (purple), and aromatic features (green).  The
silicate absorption has been applied to the PAHFIT results.  No
significant starlight is found by the PAHFIT program.  The dust
continuum is fit with up to 8 components with fixed temperatures and
is meant to produce a continuum which is physical motivated \citep[see
section 4.13 of][]{Smith07}.  This continuum fit is not unique and the
strengths of the different components do not necessarily correspond to
different real grain components.
\label{fig_pahfit_fits} }
\end{figure*}

Measuring the strengths of the aromatic features in the IRS
spectra is not straightforward given that many of 
the features are blended with each other as well as
with atomic emission lines.  Various methods have been used in the past
that attempt to define continuum points between features and measure
consistently portions of the aromatic features, but these are
prone to systematic errors as the level of blending changes with the
strength of the features.  Fortunately, \citet{Smith07} provided a
decomposition tool called PAHFIT to fit simultaneously the aromatic features,
atomic and molecular emission lines, dust and stellar continuum, and the dust
extinction in IRS spectra.  Some previous work in this area has used
similar fitting techniques \citep[e.g.,][]{Madden06}, but PAHFIT is
the first such tool to fit simultaneously all the components of
the mid-IR spectra of star forming region.  As PAHFIT takes
into account the 
blending of the aromatic and atomic features, it produces fairly precise
measurements of their individual strengths.  We used the PAHFIT program to fit
all of our M101 spectra and the resulting fits for two of the regions are shown in
Fig.~\ref{fig_pahfit_fits}.  The atomic emission line strengths from
the fits are given in Table~\ref{tab_atomic_strength}.  The aromatic
emission line strengths and equivalent widths are given in
Tables~\ref{tab_aromatic_strength} and \ref{tab_aromatic_eqw},
respectively.  The line strengths are computed by integrating the fit
profile for the specific line and does not include any contribution to
the continuum.  The equivalent width is
\begin{equation}
\int \left( \frac{I(\nu) - I(\nu)_{\mathrm{cont}}}{I(\nu)_{\mathrm{cont}}} 
  \right) \delta \lambda
\end{equation}
where $I(\nu) = I(\nu)_{\mathrm{line}} + I(\nu)_{\mathrm{cont}}$ (both
from the PAHFIT results) and
is defined such that the equivalent width is positive for an emission
line.  The uncertainties quoted are returned from PAHFIT and are the
formal uncertainties on the quantities.  These PAHFIT returned
uncertainties are based on the input spectrum uncertainties which in
turn are defined using the background noise.

There are a number of diagnostics of the physical conditions in the
HII regions contained in the IRS spectra themselves as well as
literature optical spectral studies.  The IRS spectra give probes of
the ionization ([NeIII]/[NeII], [SIV]/[SIII],
[ArIII]/[ArII]) and electron density ([SIII] 18/[SIII] 33)
\citep{Giveon02}.  The ionization probe is related to the
radiation field hardness, but it is not easy to make a direct
translation.  The optical studies provide high quality
metallicities [log(O/H)+12], electron temperatures, and ionization
(e.g., [OIII] 5007/[OII] 3727). 
\citep{Kennicutt03, Bresolin06}.  Finally, the MIPS 24~\micron\ flux
provides a rough probe of the radiation field density convolved with
the amount of dust present.  \citet{Wu06} used a similar measurement,
$L_{22 \micron}/V$, where $L_{22 \micron}$ was determined from IRS Red
Peakup images and $V$ is the volume of the region probed (in this case
the whole galaxy).  This was
described as a probe of the UV radiation field density assuming dust
absorbs most of the energy and there is a fairly constant conversion
from $L_{22 \micron}$ and total IR flux.  In our case, the $V$ is
similar for all our HII regions because we are probing regions in the
same galaxy 
with photometry performed with the same sized apertures.

\begin{figure*}[tbp]
\epsscale{1.1}
\plotone{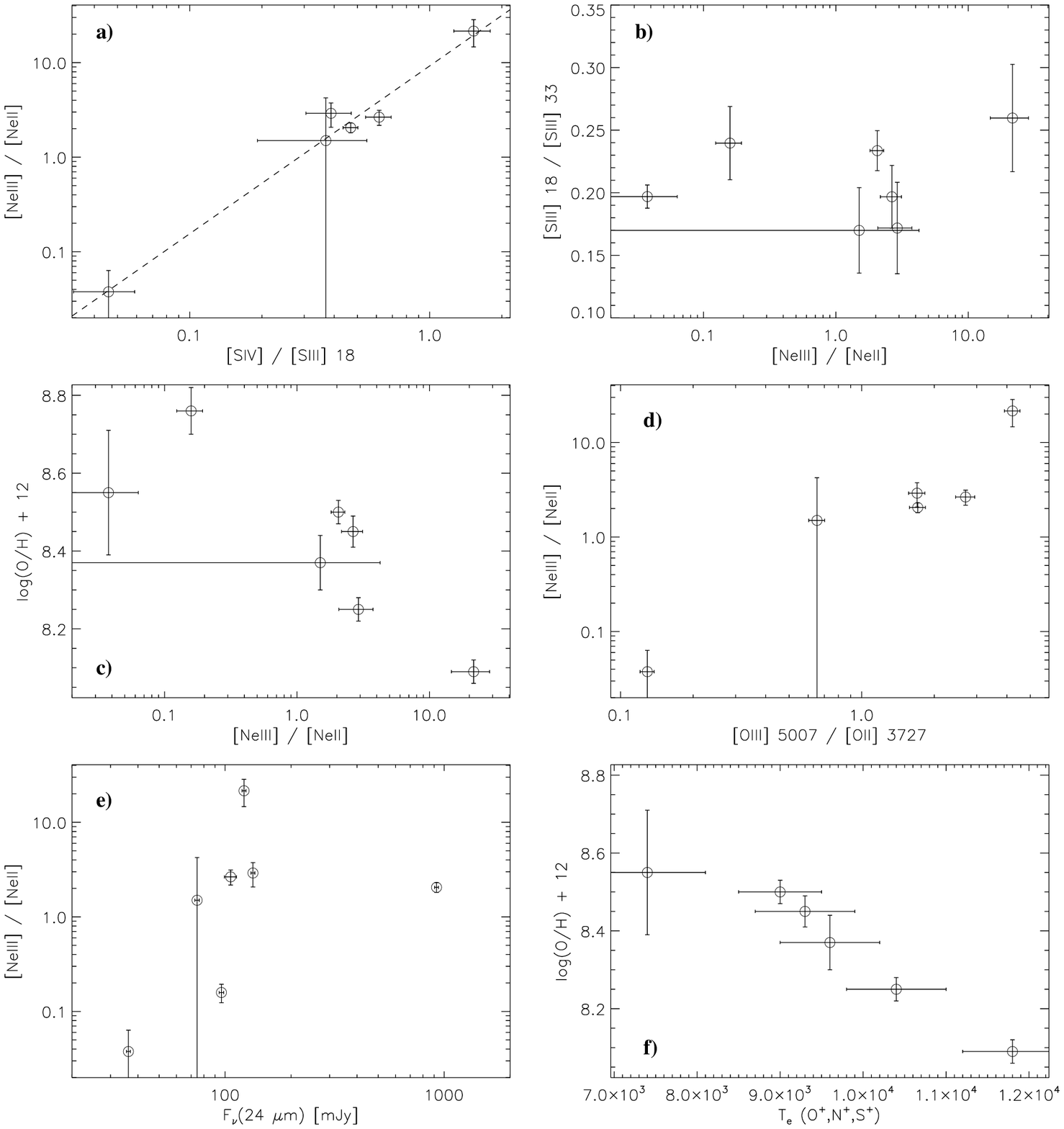}
\caption{Diagnostics of the physical conditions in HII regions are
plotted versus each other to illustrate their interdependencies.  These
probes are the ionization ([NeIII]/[NeII] and
[SIV]/[SIII] 18, dereddened [OIII] 5007/[OII] 3727), electron density
([SIII] 18/[SIII] 33), metallicity [log(O/H) + 12], the optical
emission line electron temperature [T$_e$(O$^+$,N$^+$,S$^+$)], and
radiation field intensity (MIPS 24~\micron\ flux).
\label{fig_spec_diag} }
\end{figure*}

The different diagnostics are not all independent and
Fig.~\ref{fig_spec_diag} illustrates this point.  The correlations
among the different ionization probes were seen in ISO
observations of HII regions \citep{Giveon02} and starburst galaxies
\citep{Verma03}.  Not surprisingly, we find the same for M101 HII
regions for the correlation between [NeIII]/[NeII] and [SIV]/[SIII] 18
(Fig.~\ref{fig_spec_diag}a, [ArIII]/[ArII] is not plotted as both
lines are only significantly detected in one of our HII regions).
The infrared ([NeIII]/[NeII]) and optical (dereddened [OIII] 5007/[OII 3727])
probes of ionization are seen to be correlated
(Fig.~\ref{fig_spec_diag}b), but not nearly as well as the two
infrared ionization probes (Fig.~\ref{fig_spec_diag}a).  
Metallicity and
ionization are known to be roughly correlated.  This rough correlation
is caused by decreasing line blanketing and harder spectra in hot
stars as the metallicity decreases.  
In the M101 HII regions, we also see these two quantities are correlated
(Fig.~\ref{fig_spec_diag}c), but the correlation has a larger
scatter than can be accounted for by the observational uncertainties.
Fig.~\ref{fig_spec_diag}d illustrates that all the HII
regions have similar electron densities, $< 100$~cm$^{-3}$, as
determined from the [SIII] 18/[SIII] 33 line ratio \citep{Giveon02} as
well as optical spectroscopy \citep{Kennicutt03}.  
This is
likely due to the different ionization energies between [OII]
(13.6~eV) and [NeII] (21.6~eV) and [SIII] (23.3~eV).  The rough
correlation between the optical and infrared ionization measures
indicates that there may well be significant variation in the
radiation field hardness among the different HII regions.
The radiation
field intensity as probed by the MIPS 24~\micron\ flux is not
correlated with any of the physical quantities
(Fig.~\ref{fig_spec_diag}e) although this lack of correlation is
mainly based on the large 24~\micron\ flux seen for NGC~5461. 
Finally, the optically determined electron temperatures are well correlated
with metallicity (Fig.~\ref{fig_spec_diag}f).
Overall, there seem to be three 
quantities that probe different physical properties of the HII
regions.  They are log(O/H)+12 (metallicity and electron temperature),
[NeIII]/[NeII] and [SIV]/[SIII] 18 (ionization), and
MIPS 24~\micron\ flux (radiation field intensity).

Given that the [NeIII]/[NeII] and [SIV]/[SIII]~18 ratios are well
correlated, it is possible to combine the two ratios into a
composite ionization measurement with smaller
uncertainties.  In this paper and \citet{Engelbracht08SB}, we
propose a composite measure called the ionization
index (or II for short).  It is defined as the weighted average
of log([SIV]/S[III]~18) converted to the [NeIII]/[NeII] scale and
log([NeIII]/[NeII]).  The conversion of [SIV]/S[III]~18 to the
[NeIII]/[NeII] scale is done using the observed correlation for the
M101 HII regions and the starburst galaxies presented by
\citet{Engelbracht08SB}.  This correlation for just the M101 HII
regions (Fig.~\ref{fig_spec_diag}a) is
\begin{equation}
\log \left( y \right) =
    (0.93 \pm 0.11) + (1.85 \pm 0.34) \log \left( x \right)
\end{equation}
where $x =$ [SIV]/[SIII]~18 and $y =$ [NeIII]/[NeII].
The combined correlation of the M101 HII regions and the starburst
galaxies is
\begin{equation}
\log \left( y \right) =
    (0.71 \pm 0.08) + (1.58 \pm 0.13) \log \left( x \right).
\end{equation}
We adopt the combined correlation to determine the II values for the
M101 HII regions.  Thus, II can be calculated from
\begin{equation}
\mathrm{II} = \log \left( \frac{\mathrm{[NeIII]}}{\mathrm{[NeII]}}
   \right) + \left[ 0.71 + 1.58 
   \log \left( \frac{\mathrm{[SIV]}}{\mathrm{[SIII] 18}} \right) \right].
\end{equation}
We use the II values for the rest of this paper instead of the
[NeIII]/[NeII] or [SIV]/[SIII]~18 ratios as the II has less noise and
is available for all the M101 HII regions.

We have chosen to call the combined measurement of [NeIII]/[NeII] and
[SIV]/[SIII] an ionization index as this is the physical measurement
which is being made.  Often these two ratios are used as an indicator
of the radiation field hardness (or $T_{eff}$).  These ratios are
dependent on the metallicity, ionization parameter and morphology. and
age of an HII region.  The dependence is such that the ratios decrease
as the metallicity increases, the ionization parameter decreases, and
the age decreases \citep{Rigby04}.  This explains why these ratios are
only roughly correlated with metallicity.  The scatter in the
metallicity versus [NeIII]/[NeII] ratio (Fig.~\ref{fig_spec_diag}c) is
likely due to different HII region ages and/or ionization parameters.
Disentangling the different effects is complicated and beyond the
scope of this paper as we are only interested in these ratios as a
measure of the amount of processing which might have taken place in
the HII regions.

\begin{figure*}[tbp]
\epsscale{1.1}
\plottwo{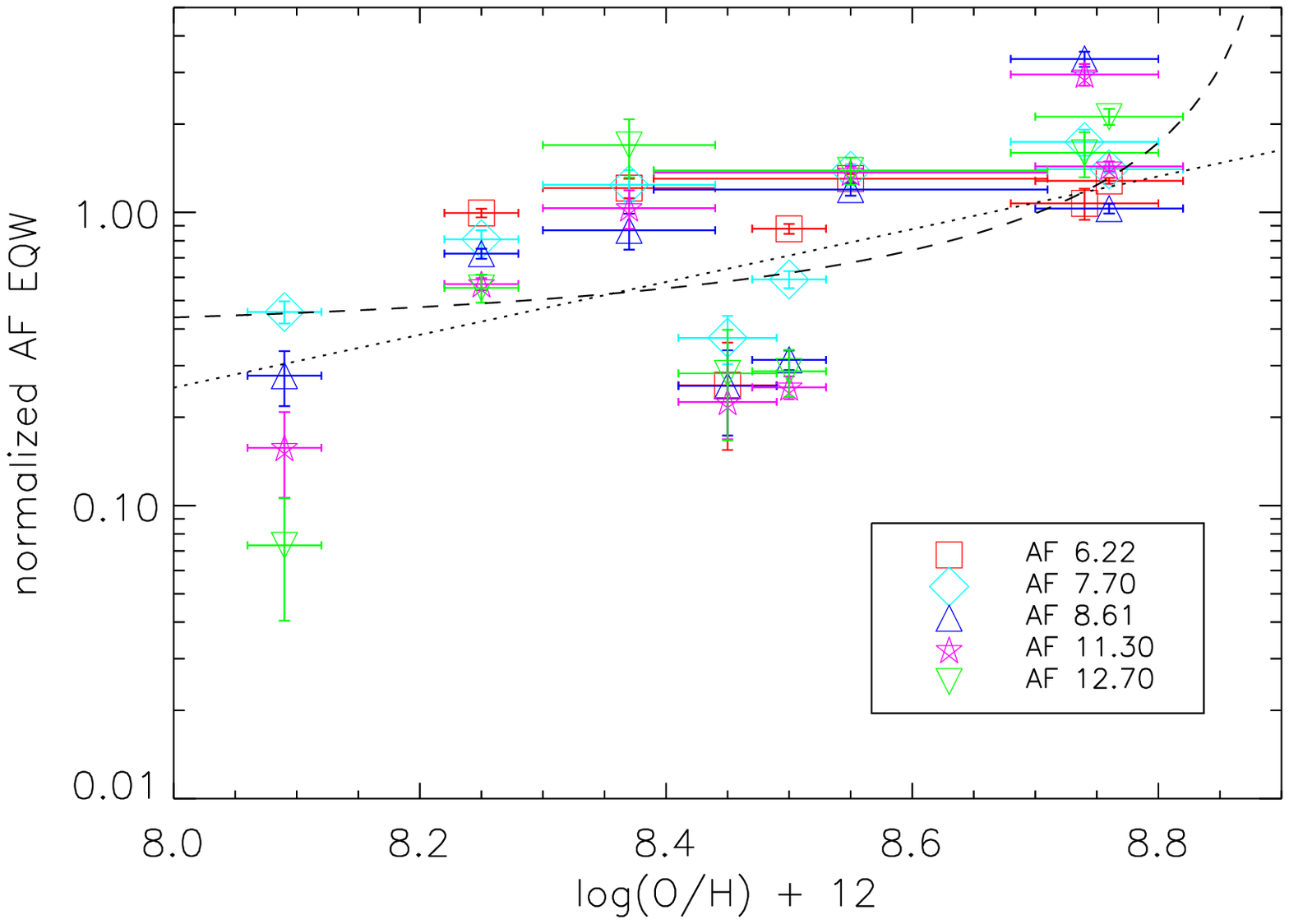}{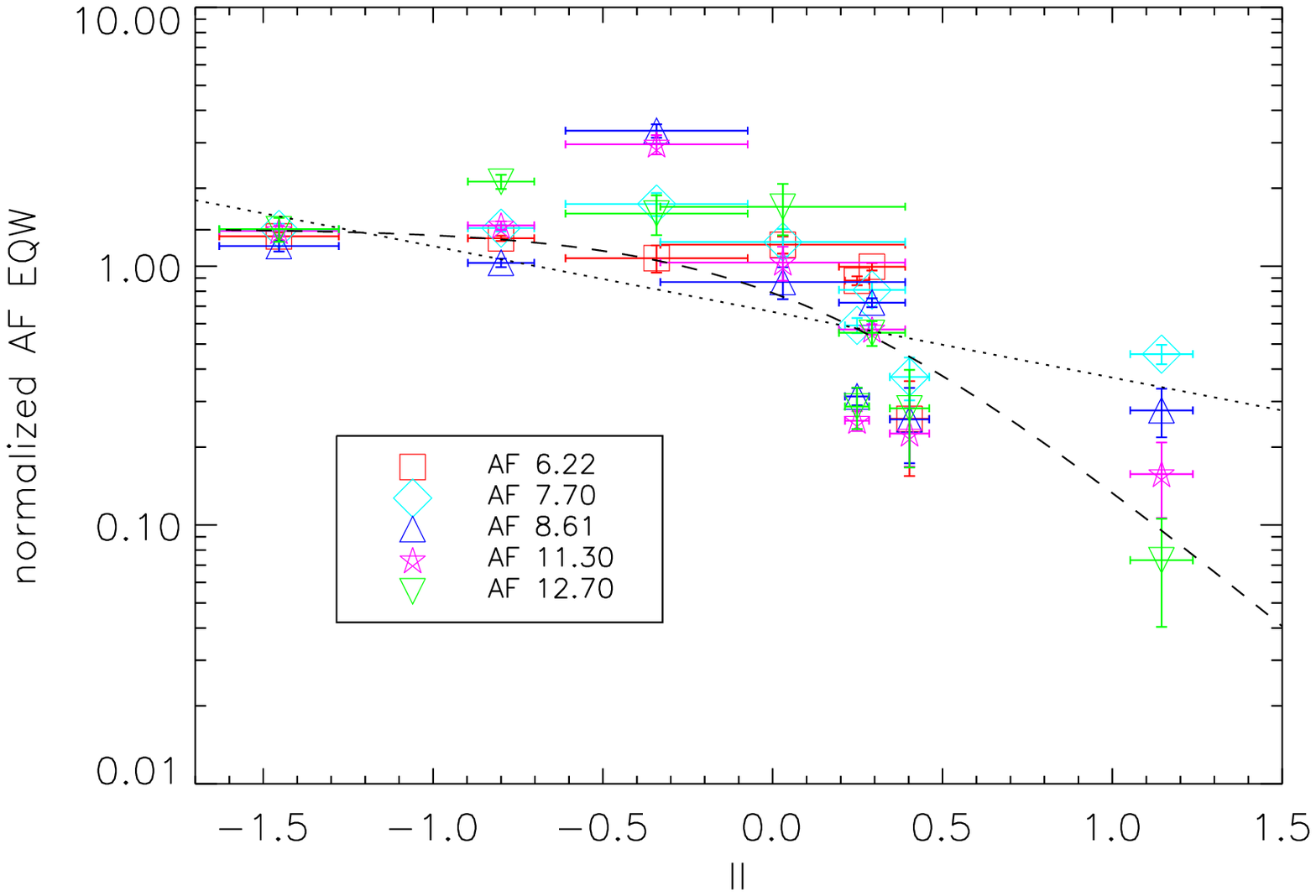}
\caption{The normalized equivalent widths of the 5 highest equivalent
width aromatic features (AFs) are plotted versus metallicity (left)
and ionization index (II [see \S\ref{sec_spec_aromatic})], right).
The normalization was done to 
the average equivalent width of each AF.  The normalization values
are 0.87, 3.01, 1.04, 1.32, and 0.56 \micron\ for the 6.22, 7.70, 8.61,
11.30, and 12.70~\micron\ aromatic features, respectively.  The dotted
and dashed lines give the best fit power law and power law plus
constant, respectively (see \S\ref{sec_spec_aromatic}).
\label{fig_spec_eqw}}
\end{figure*}

The equivalent widths of the aromatic features are plotted versus
metallicity and II in Fig.~\ref{fig_spec_eqw}.  The equivalent width
measures the strength of the aromatic features versus the underlying
mid-infrared dust continuum emission.  As both the aromatic features
and underlying continuum are measured at the same wavelengths, they
are likely to have undergone similar excitation and, therefore, the
equivalent width is a measurement of the abundance ratio of the
aromatic carriers to small dust grains.  Visually, it appears that 
the equivalent width of the aromatic features is better correlated with
II than with metallicity.  We have quantified this by fitting the
measurements with two different functions.  The first is a simple
power law
\begin{equation}
y = ax^b
\end{equation}
where $x = 10^{\mathrm{II}}$ and y is the normalized aromatic equivalent width.
The second equation is a power law plus a constant with the form
\begin{equation}
y = \left[ ax^b + c \right]^{-1}.
\label{eq_powerlaw_const}
\end{equation}
The best fits using both of these equations are shown in
Fig.~\ref{fig_spec_eqw}.  The reduced $\chi^2$ values for the data
versus metallicity are 356 and 339 for the simple power and power law
plus a constant, respectively.  The reduced $\chi^2$ values for the
data versus II are 240 and 195 for the simple power and power law
plus a constant, respectively.  This indicates that the
aromatic feature equivalent widths are better correlated with ionization than
metallicity.  In addition, the functional form favored 
is the power law plus constant (Eq.~\ref{eq_powerlaw_const}).  The
behavior of the aromatic equivalent widths can be described as
constant up to a threshold II ($\sim 0$ which is a [NeIII]/[NeII]
value of 1) and then a reduction in equivalent width with a power law form at
higher II values.  The best fit between the II and normalized
equivalent width measurements is
\begin{equation}
y = \left[ 0.55x^{0.98} + 0.73 \right]^{-1}.
\end{equation}
The expected values for the equivalent widths for a particular
aromatic feature can be determined from this equation multiplied by the
average equivalent widths that are given in the caption of
Fig.~\ref{fig_spec_eqw}. 

\begin{figure*}[tbp]
\epsscale{1.1}
\plottwo{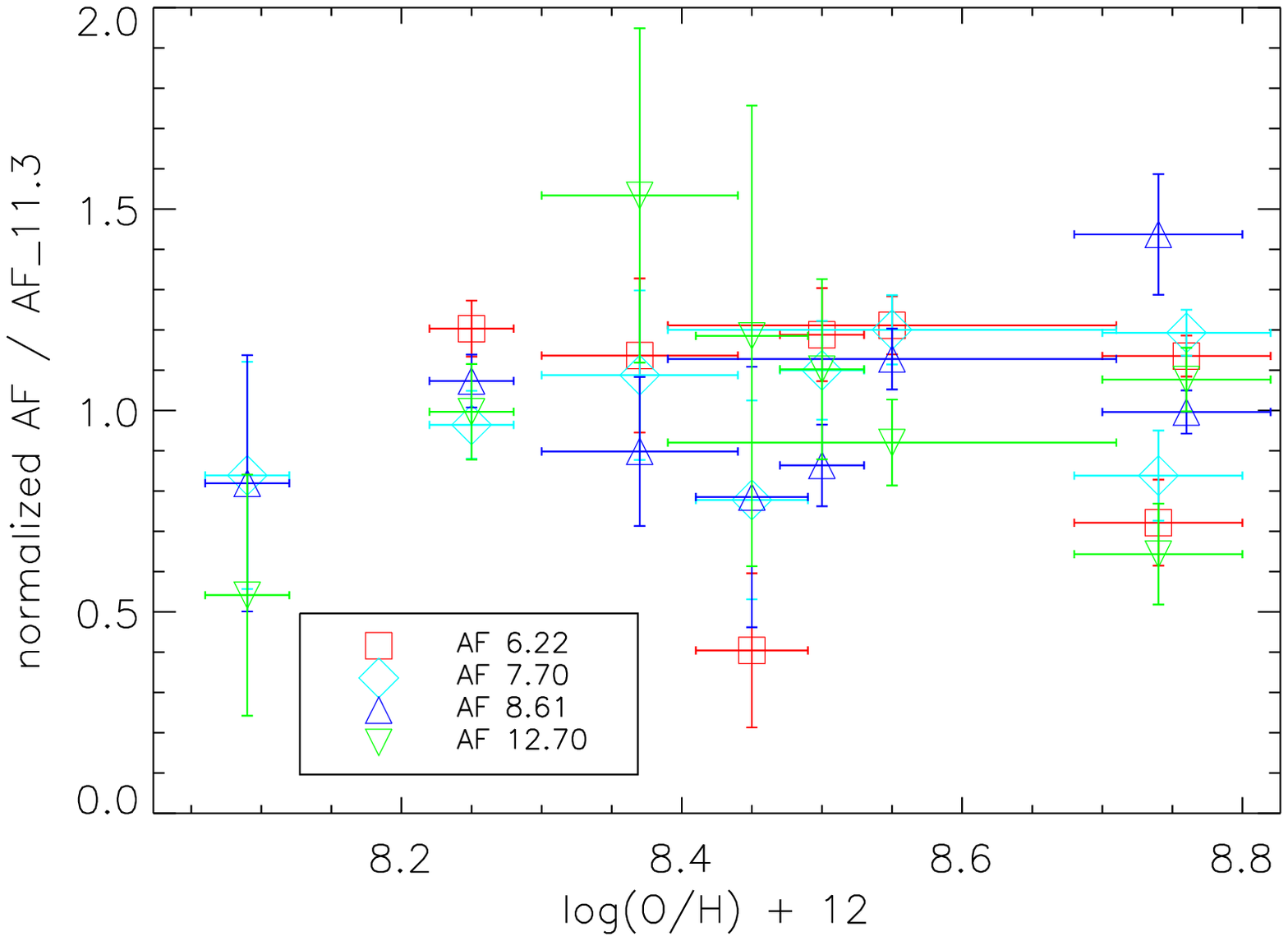}{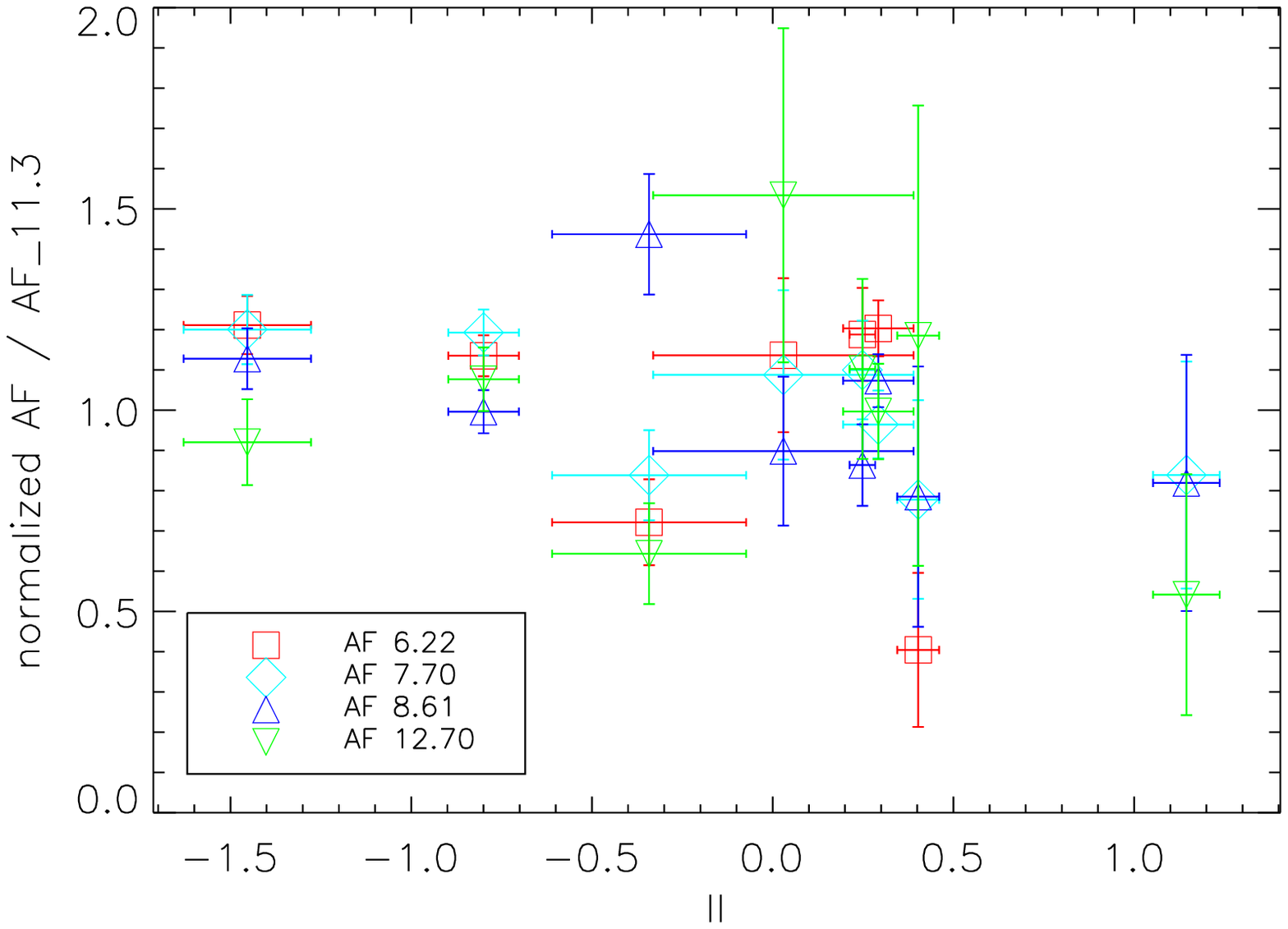}
\caption{The ratio of the strength of the 4 strongest
aromatic features (AFs) to the 11.3 AF are plotted versus metallicity (left)
and ionization index (II, right).  The normalization was done to
the average ratio of each AF.  The normalization values are 1.01,
3.41, 0.80, and 0.57 for the 6.22, 7.70, 8.61, 
and 12.70~\micron\ aromatic features, respectively.
\label{fig_spec_ratio}}
\end{figure*}

The ratio of the integrated strengths of different aromatic features
is a diagnostic of the physical mechanism of processing.  For example,
models using PAHs as the carriers of the 
aromatic features predict the strength of the 6.22, 7.70, and
8.61~\micron\ features to strengthen relative to the 11.3 and 12.7~\micron\
features as the ionization level increases or to weaken as the smaller
PAHs are removed \citep{Bakes01, Peeters02}.  In
Fig.~\ref{fig_spec_ratio} we show the ratio of the 4 strongest
aromatic features to the 11.3~\micron\ feature versus metallicity and
II.  The ratios do not vary significantly from object to object nor
does there seem to be a significant trend with either of the two
diagnostics.  While there is a large scatter in the ratios of around a
factor of two, the large measurement uncertainties mean that the
ratios could be consistent with no change with either of the two
diagnostics.  This is different than the preliminary result reported
by \citet{Gordon06IRDiag} where significant variation in the ratios
was found.  We have improved both the IRS data reductions and
aromatic feature equivalent width measurements, which accounts for the
differences.  The lack of ratio variations is interesting in light of
the strong weakening of the aromatics above a threshold in II
(Fig.~\ref{fig_spec_eqw}).  The aromatic features weaken by over a
factor of 10, but the ratio of different aromatics does not change by
more than a factor of 2.

\subsection{Photometric Measures of Aromatics}
\label{sec_photo_hII}

While IRS spectroscopy is clearly the most accurate way to study the
aromatic features, it would be useful to be able to probe the aromatic
feature strengths using some combination of IRAC and MIPS imaging.
Having a photometric probe to the aromatic features would allow for a
larger range of objects and environments to be studied given that
imaging is faster to acquire and can probe regions that are too faint
for spectroscopy.  \citet{Engelbracht05SB} proposed the combination of
8-to-4.5 and 8-to-24~\micron\ flux density ratios as just such a
measure of the aromatic feature strength given that the IRAC
8~\micron\ band probes the strong 8~\micron\ aromatic complex
(composed of the 7.7, 8.3, and 8.6~\micron\ aromatic features) and the
IRAC 4.5~\micron\ and MIPS 24~\micron\ bands probe different parts of the small grain
dust emission.

\begin{figure}[tbp]
\epsscale{1.2}
\plotone{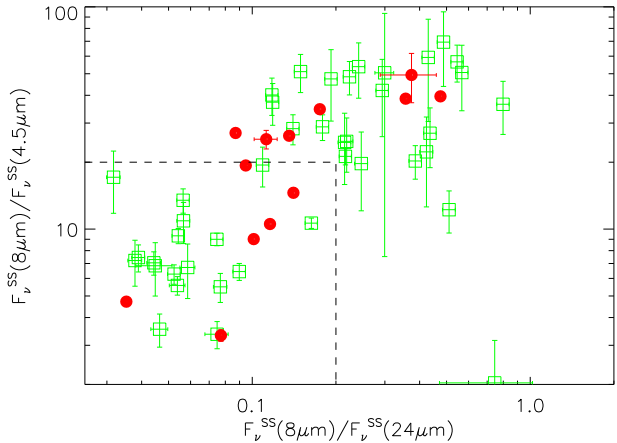}
\caption{The $f_{\nu}^{SS}(8~\micron)/f_{\nu}^{SS}(4.5~\micron)$ ratio
is plotted versus the
$f_{\nu}^{SS}(8~\micron)/f_{\nu}^{SS}(24~\micron)$ ratio where the
superscript ``SS'' designates these fluxes as stellar subtracted.  The
contributions to the measured fluxes from stellar emission was
subtracted by scaling the measured 3.6~\micron\ flux by 0.57, 0.23,
and 0.031 for the 4.5~\micron, 8.0~\micron, and 24~\micron\ bands,
respectively.  The HII regions are plotted as solid, red circles and
the starburst galaxies \citep{Engelbracht08SB} as open,
green squares.  The error bars are derived from the measurement
uncertainties and include the contribution from stellar subtraction.
The dotted line shows a spectroscopically-chosen color separation
between galaxies with (outside the box) and without (inside the box)
aromatic emission \citep{Engelbracht08SB}.
\label{fig_photo_plot} }
\end{figure}

The 8-to-4.5 and 8-to-24~\micron\ flux density ratios are
measuring something very 
similar to an aromatic feature equivalent width as it can be thought
of as the ratio of the aromatic emission divided by the underlying
small grain emission.  It is not the same as an equivalent width as
the small grain emission measurement is done at shorter and longer
wavelengths (4.5 and 24~\micron, respectively, instead of 8~\micron)
than is done when using the IRS 
spectroscopy.  In other words, the 4.5~\micron\ flux probes
hotter and 24~\micron\ cooler dust grains 
than the 8~\micron\ spectroscopic continuum.  Thus, variations in
8-to-4.5 and 8-to-24~\micron\ can be due to changes in the dust
emission continuum instead of the 8~\micron\ aromatic complex.  Yet if
both ratios change in a consistent manner, it is highly likely it is
the 8~\micron\ aromatic complex which is varying.
In Figure~\ref{fig_photo_plot} we plot the
8-to-24~\micron\ versus 
8-to-4.5~\micron\ flux density ratios for a sample of M101 HII regions
and a sample of starburst galaxies \citet{Engelbracht08SB}.  The M101
HII region photometry was performed on the MIPS 24~\micron\
and IRAC 3.6, 4.5, \& 8.0 images convolved to the 24~\micron\ PSF
(\S\ref{sec_res_match}).  The photometry was measured using a
10\arcsec\ radius object aperture and a background annulus with radii
from 13 to 20\arcsec and is given in 
Table~\ref{tab_m101_regphot}.  The IRAC 3.6~\micron\ band is dominated by
stellar emission and was used to subtract the stellar contribution
from the longer wavelength measurements \citep{Engelbracht05SB}.  The
starburst photometry is from \citet{Engelbracht08SB} and was also
corrected for stellar emission.

\begin{deluxetable*}{lrrrrr}
\tablewidth{0pt}
\tablecaption{Photometry for M101 regions \label{tab_m101_regphot}}
\tablehead{ \colhead{name} & \colhead{IRAC1} & \colhead{IRAC2} & \colhead{IRAC3} & \colhead{IRAC4} & \colhead{MIPS24} \\
 & \colhead{3.6~\micron} & \colhead{4.5~\micron} & \colhead{5.8~\micron} & \colhead{8.0~\micron} & \colhead{24~\micron} \\
 & \colhead{[mJy]} & \colhead{[mJy]} & \colhead{[mJy]} & \colhead{[mJy]} & \colhead{[mJy]}}
\startdata
   Nucleus  & $  15.00 \pm   0.119$ & $   9.70 \pm   0.083$ & $  19.20 \pm   0.222$ & $  48.90 \pm   0.675$ & $  96.20 \pm   1.221$ \\
  Hodge~602  & $   3.09 \pm   0.595$ & $   1.94 \pm   0.364$ & $   3.78 \pm   0.662$ & $   9.35 \pm   1.545$ & $  23.20 \pm   3.654$ \\
 Hodge~1013  & $   1.59 \pm   0.051$ & $   1.47 \pm   0.035$ & $   7.33 \pm   0.166$ & $  20.00 \pm   0.447$ & $ 112.00 \pm   0.531$ \\
   Searle~5  & $   1.17 \pm   0.017$ & $   1.00 \pm   0.012$ & $   4.82 \pm   0.039$ & $  13.20 \pm   0.126$ & $  36.20 \pm   0.128$ \\
   NGC~5461  & $   7.01 \pm   0.143$ & $   6.96 \pm   0.121$ & $  28.40 \pm   0.534$ & $  82.00 \pm   1.392$ & $ 923.00 \pm   5.994$ \\
   NGC~5462  & $   1.37 \pm   0.047$ & $   1.16 \pm   0.040$ & $   4.08 \pm   0.086$ & $  10.40 \pm   0.213$ & $  74.50 \pm   0.884$ \\
   NGC~5455  & $   1.92 \pm   0.009$ & $   1.75 \pm   0.009$ & $   5.36 \pm   0.031$ & $  13.10 \pm   0.082$ & $ 134.00 \pm   0.388$ \\
   NGC~5447  & $   1.75 \pm   0.104$ & $   1.47 \pm   0.091$ & $   4.62 \pm   0.342$ & $  12.30 \pm   0.898$ & $ 106.00 \pm   6.463$ \\
   Hodge~67  & $   0.19 \pm   0.008$ & $   0.17 \pm   0.006$ & $   0.43 \pm   0.012$ & $   1.02 \pm   0.026$ & $   6.94 \pm   0.071$ \\
Hodge~70/71  & $   0.26 \pm   0.008$ & $   0.26 \pm   0.006$ & $   0.52 \pm   0.017$ & $   1.23 \pm   0.043$ & $  10.20 \pm   0.120$ \\
  Searle~12  & $   0.45 \pm   0.007$ & $   0.44 \pm   0.005$ & $   0.79 \pm   0.009$ & $   1.76 \pm   0.019$ & $  16.30 \pm   0.079$ \\
   NGC~5471  & $   1.56 \pm   0.012$ & $   1.80 \pm   0.011$ & $   2.10 \pm   0.019$ & $   4.66 \pm   0.035$ & $ 122.00 \pm   0.777$ \\
  Hodge~681  & $   0.12 \pm   0.008$ & $   0.10 \pm   0.005$ & $   0.11 \pm   0.005$ & $   0.14 \pm   0.004$ & $   1.53 \pm   0.042$ \\
\enddata
\end{deluxetable*}

The plot shown in Fig.~\ref{fig_photo_plot} is similar to Fig.~1 of
\citet{Engelbracht05SB} except that we have plotted the
8-to-4.5~\micron\ instead of 4.5-to-8~\micron\ flux ratio so that both
the ratios plotted behave in a similar sense where larger ratios imply
stronger aromatic features.  In addition, we have subtracted the
stellar contribution from all three bands involved using scaled
3.6~\micron\ flux densities.  These two ratios probe the behavior of
the 8~\micron\ aromatic complex plus underlying dust continuum
emission versus shorter wavelength dust continuum emission (8.0/4.5)
and longer wavelength dust continuum emission (8/24).
Fig.~\ref{fig_photo_plot} shows that the M101 HII regions and
starburst galaxies show the same behavior and cover similar ranges in
both ratios.  This is not surprising given that the simplest model of
a starburst is a collection of HII regions.  However, it is important to
confirm that HII regions have mid-IR SEDs similar to those of
starburst galaxies and, as a result, are reasonable analogs for
studying dust emission from starburst galaxies.

\begin{figure*}[tbp]
\epsscale{1.1}
\plottwo{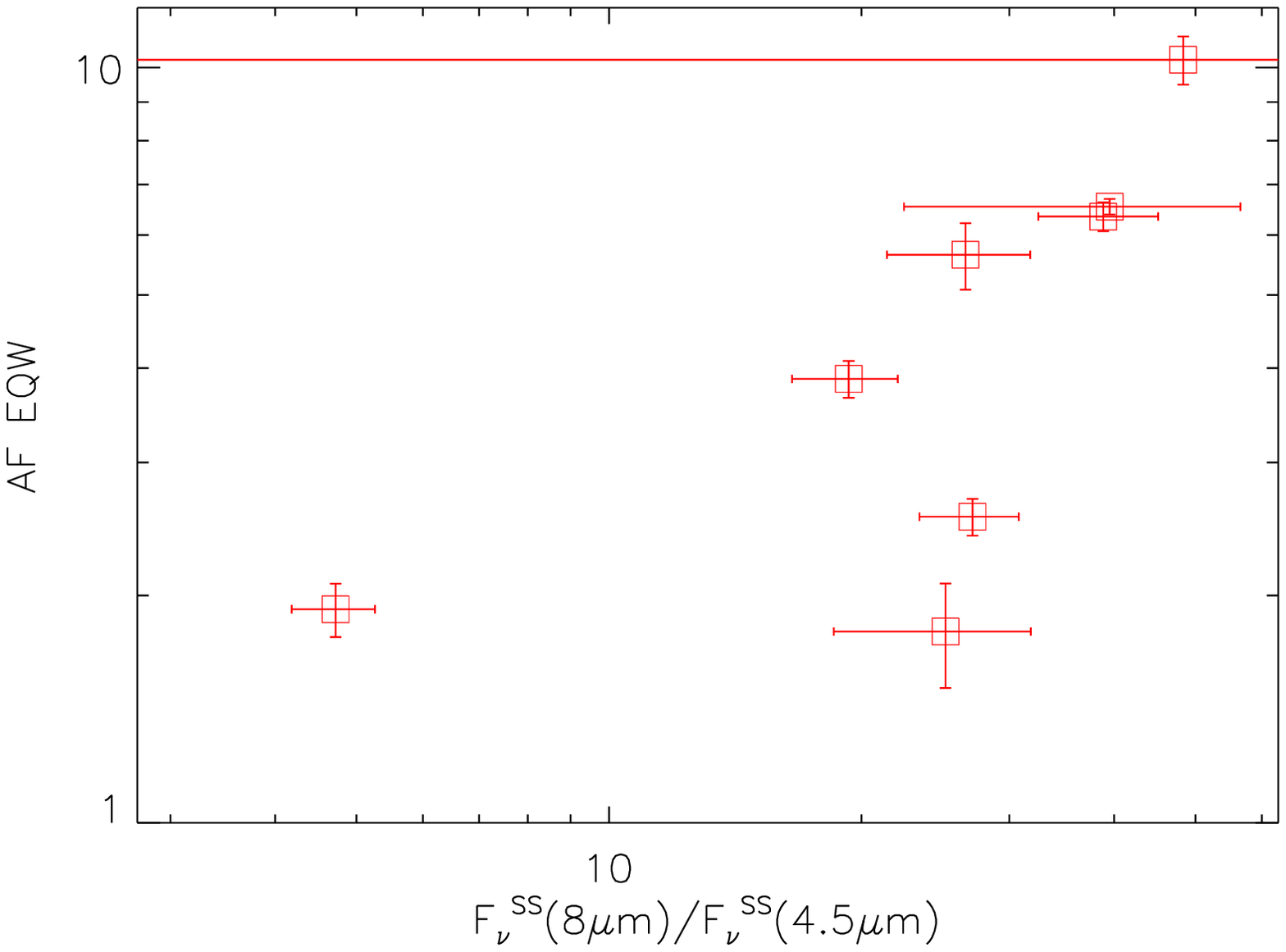}{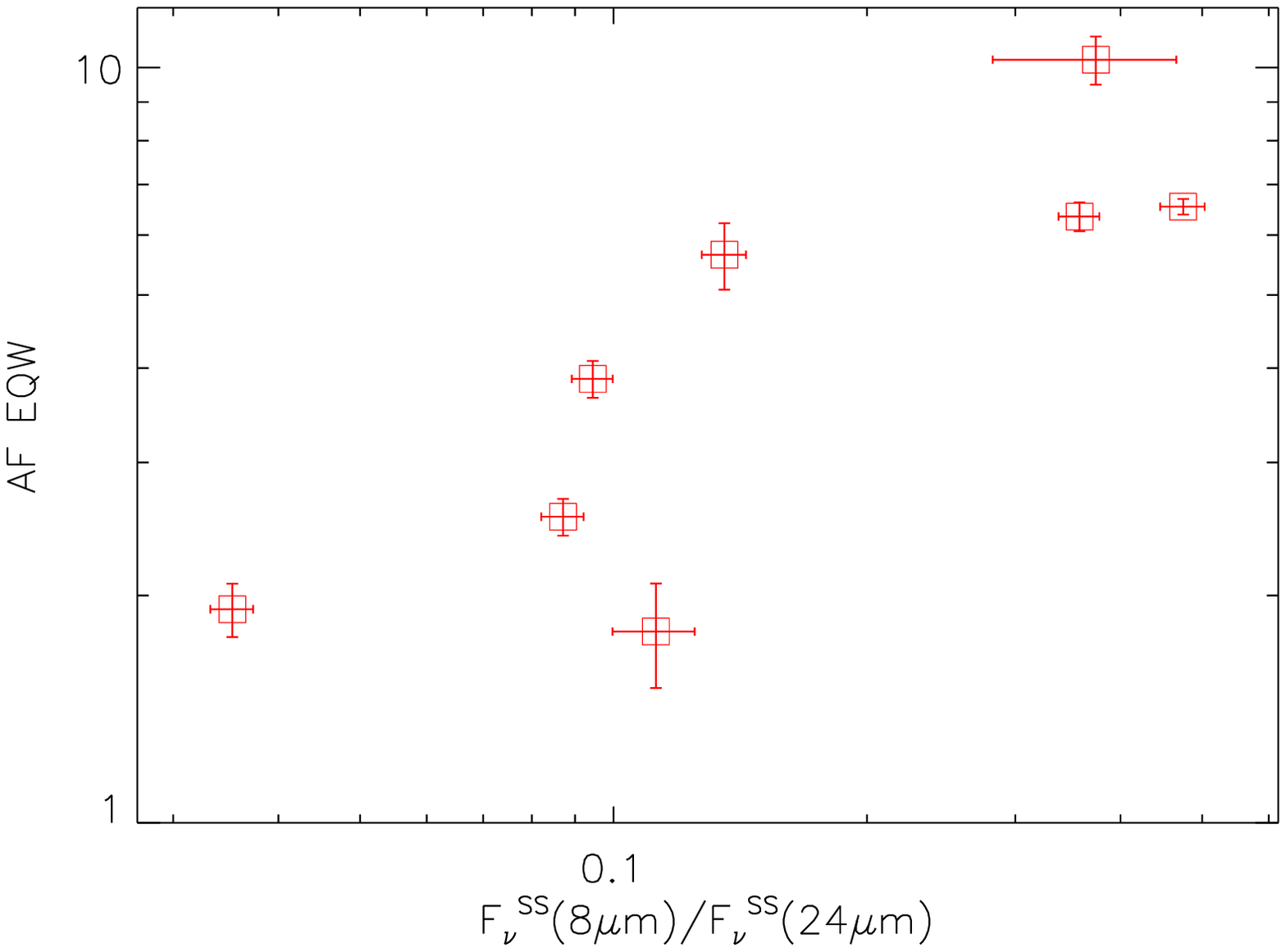}
\caption{The $F_{\nu}^{SS}(8~\micron)/F_{\nu}^{SS}(4.5~\micron)$ and
$F_{\nu}^{SS}(8~\micron)/F_{\nu}^{SS}(24~\micron)$ ratios are shown
plotted versus measured 8~\micron\ aromatic feature complex equivalent
width. \label{fig_ratio_spectro} }
\end{figure*}

Given that we have spectroscopic and photometric measurements of
the 8~\micron\ aromatic feature complex, we can see how the
photometric measure compares with the more accurate spectroscopic
measurement.  Figure~\ref{fig_ratio_spectro} gives this comparison
where the measured spectroscopic equivalent width of the 8~\micron\
aromatic complex is plotted versus the 8-to-4.5~\micron\ and
8-to-24~\micron\ flux density ratios.  It is clear that these two flux
ratios are rough measures of the 8~\micron\ complex.  The
8-to-24~\micron\ ratio is probably a better measure given that the
relationship is more linear than that using the 8-to-4.5~\micron\
ratio and the measurement uncertainties are smaller.

\begin{figure}[tbp]
\epsscale{1.2}
\plotone{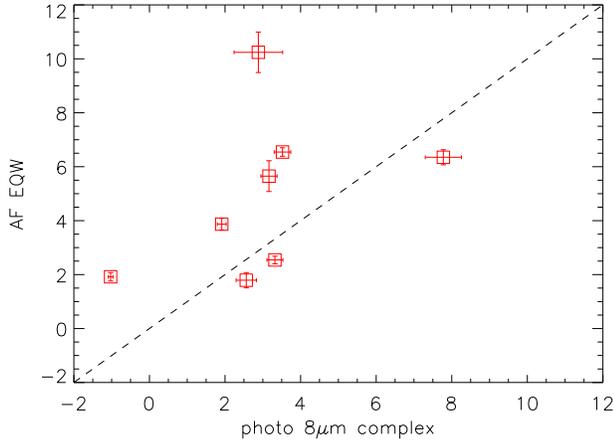}
\caption{The photometric (see \S\ref{sec_photo_hII}) versus
spectroscopic measurements of the 
8~\micron\ aromatic complex equivalent width are plotted.  The dashed
line shows the one-to-one relationship.
\label{fig_photo_spectro} }
\end{figure}

It is possible
to combine both ratios in a way to create a better measure of the
8~\micron\ complex.  The basic idea is to use the 4.5 and 24~\micron\
measurements to predict the continuum at 8~\micron\ due to small grain
emission and combine this continuum prediction with the measured
8~\micron\ flux to create a photometric measure of the 8~\micron\
equivalent width.  Following \citet{Engelbracht08SB}, we find that it
is best predicted using  
\begin{equation}
F_{\nu}(8\micron)_{\mathrm{cont}} = F_{\nu}(4.5\micron)^{0.66} \times F_{\nu}(24\micron)^{0.34}
\end{equation}
which results from assuming the small grain emission follows a power
law.  Unlike \citet{Engelbracht08SB}, we have not used the stellar
subtracted fluxes to determine $F_{\nu}(8~\micron)_{\mathrm{cont}}$ as
we do not have K band fluxes to help measure the stellar flux and the
stellar contribution is expected to be small for these resolved HII
regions.  Using the stellar subtracted fluxes produces systematically
larger photometric than the measured spectroscopic equivalent widths.
The photometric equivalent width is then
\begin{equation}
\mathrm{EQW}(8~\micron)_{\mathrm{photo}} = \left(
   \frac{F_{\nu}(8\micron)}{F_{\nu}(8\micron)_{\mathrm{cont}}} - 1
\right) \Delta\lambda.
\end{equation}
where $\delta\lambda = 2.69$~\micron\ and is the width of the IRAC
8~\micron\ band.  This assumes that the 8~\micron\ complex fills the
IRAC 8~\micron\ band which is the case for many of the HII regions,
but not all of them (see Fig.~\ref{fig_m101_spec}).
We have computed the photometry equivalent widths for the M101 regions
and plotted them versus the spectroscopic equivalent widths in
Fig.~\ref{fig_photo_spectro}.  The photometric equivalent width
measurement produces a good rough measurement of the 8~\micron\
aromatic complex for the M101 regions.  A similar result is found by
\citet{Engelbracht08SB} for the starburst galaxy sample. 

\subsection{Morphology of Aromatic Emission}
\label{sec_morphology}

\begin{figure*}[tbp]
\epsscale{1.0}
\plotone{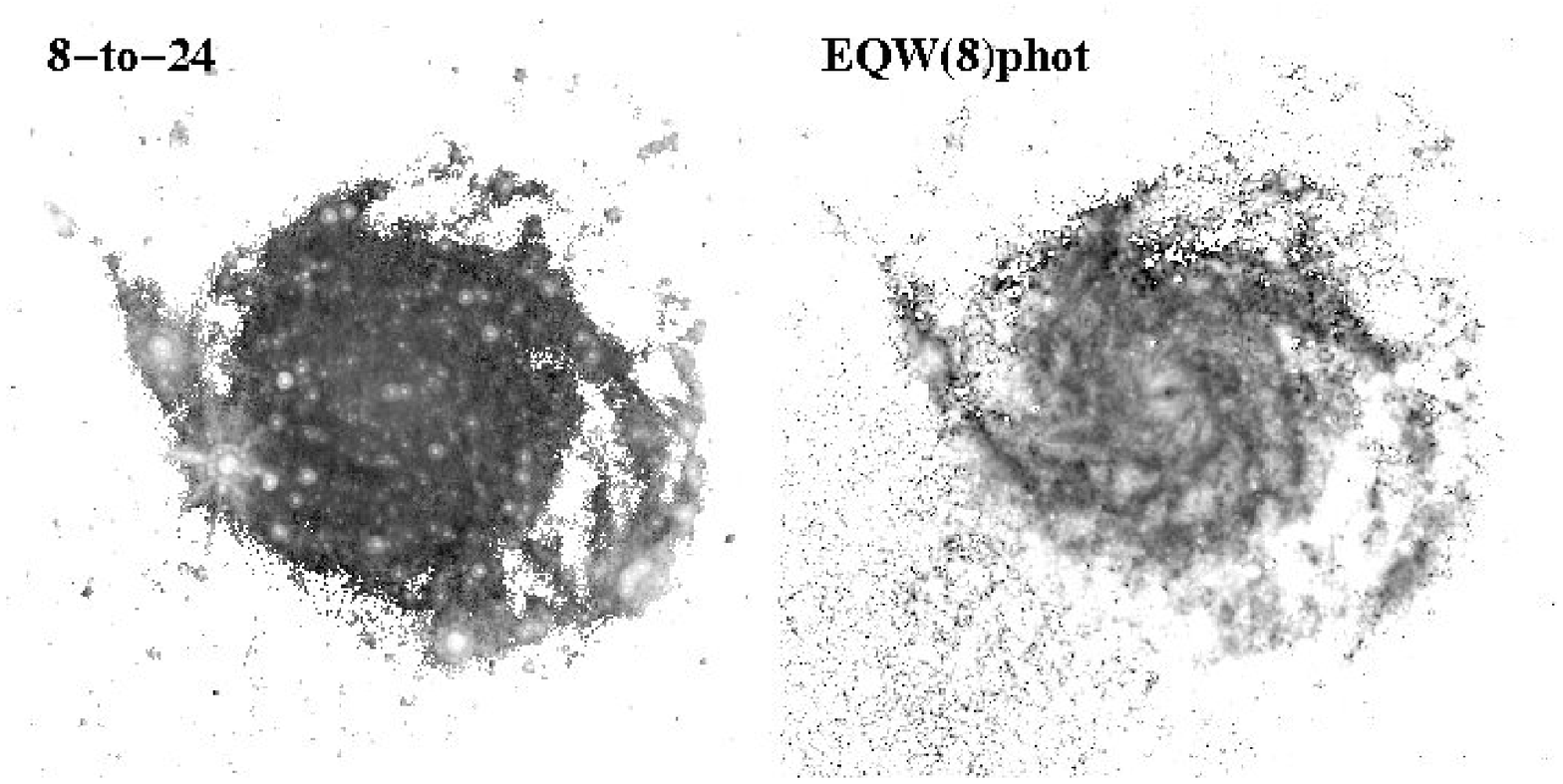}
\caption{The 8-to-24 and $\mathrm{EQW}(8~\micron)_{\mathrm{photo}}$
images of M101 are shown.  The images are displayed in reverse B/W
using a asinh stretch with ranges from 0.1--2.0 and 3--30 and pivots
of 0.5 and 5.0, respectively.
The inner 20$\arcmin$ of M101 is shown.
Only pixels with measurements above the sky noise are displayed.  The
$\mathrm{EQW}(8~\micron)_{\mathrm{photo}}$ image shows significant
noise due to instrumental signatures present in the IRAC 8~\micron\
mosaic (see Fig.~\ref{fig_m101_fullres}).
\label{fig_photo_image} }
\end{figure*}

\begin{figure*}[tbp]
\epsscale{1.0}
\plotone{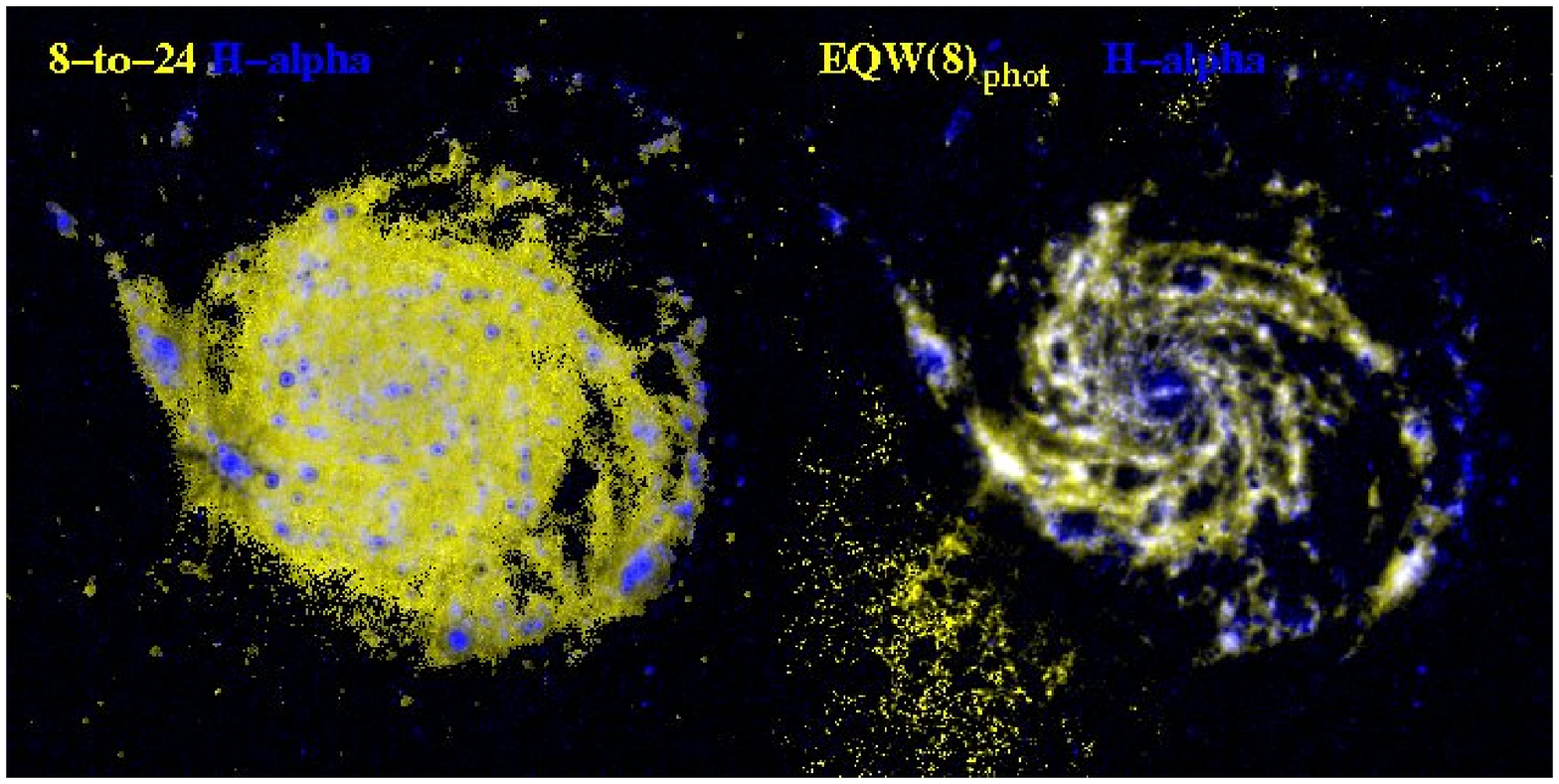}
\caption{The 8-to-24 and $\mathrm{EQW}(8~\micron)_{\mathrm{photo}}$
images of M101 are shown in yellow along with the H$\alpha$ image in
blue.  The inner 20$\arcmin$ of M101 is shown.
Only pixels with measurements above the sky noise are
displayed.
\label{fig_photo_halpha} }
\end{figure*}

The IRAC and MIPS images of M101 can be used to probe the morphology
of M101 in the 8~\micron\ aromatic complex by applying the photometric
measurement techniques introduced in \S\ref{sec_photo_hII}.
Fig.~\ref{fig_photo_image} shows the simple 8-to-24~\micron\ flux
density ratio image and the $\mathrm{EQW}(8~\micron)_{\mathrm{photo}}$
image.  The IRAC images were convolved to the 24~\micron\ resolution
(\S\ref{sec_res_match}) of 6$\arcsec$ (0.2~kpc at 6.7~Mpc) before
computing either composite image.  Caution should be exercised in
interpreting these images as both involve stellar subtracted images.
The stellar subtraction is done using a multiplicative scaling of the
IRAC 3.6~\micron\ image and the accuracy of this method has only been
shown for integrated fluxes.  This method will oversubtract regions
where the emission in the 3.6~\micron\ image is dominated by dust
emission instead of stellar emission.  While the
$\mathrm{EQW}(8~\micron)_{\mathrm{photo}}$ image is expected to be a
more accurate estimate of the true 8~\micron\ aromatic complex
equivalent width than the 8-to-24~\micron\ image, it suffers from
higher noise given the sensitivity of this measurement to the more
uncertain stellar subtracted 4.5~\micron\ flux and 8~\micron\
instrumental residuals.  Both images show that the aromatic features
are depressed in HII regions (white regions in
Fig.~\ref{fig_m101_fullres}).  This is illustrated in
Fig.~\ref{fig_photo_halpha} where the H$\alpha$ image is shown in blue
and the 8-to-24~\micron\ and
$\mathrm{EQW}(8~\micron)_{\mathrm{photo}}$ images are shown in yellow.
The preponderance of yellow and blue in the first image shows that the
8-to-24~\micron\ is anti-correlated with H$\alpha$ emission.  This is
an indication that the 8~\micron\ flux is depressed and/or the
24~\micron\ flux is enhanced in HII regions.  The
$\mathrm{EQW}(8~\micron)_{\mathrm{photo}}$+H$\alpha$ image directly
probes the strength of the 8~\micron\ aromatic complex versus
H$\alpha$ and shows that the separation between the yellow and blue is
not as clean as in the 8-to-24~\micron+H$\alpha$ image.  Thus, the
enhancement of 24~\micron\ emission in HII regions is a significant
contributer to the morphology of the 8-to-24~\micron\ image.  In other
words, the dust in HII regions is hotter (emitting more at 24~\micron)
than the surrounding dust, but this hotter dust does not have a
corresponding increase in the 8~\micron\ continuum or aromatic
emission to keep the 8-to-24~\micron\ ratio constant.  But there is
clearly also a change in the 8~\micron\ emission, given that there are
large HII complexes (e.g., NGC 5462, NGC 5447) in the second image
which show the expected morphology with aromatic emission (yellow)
surrounding the H$\alpha$ emission.

\begin{figure*}[tbp]
\epsscale{1.0}
\plotone{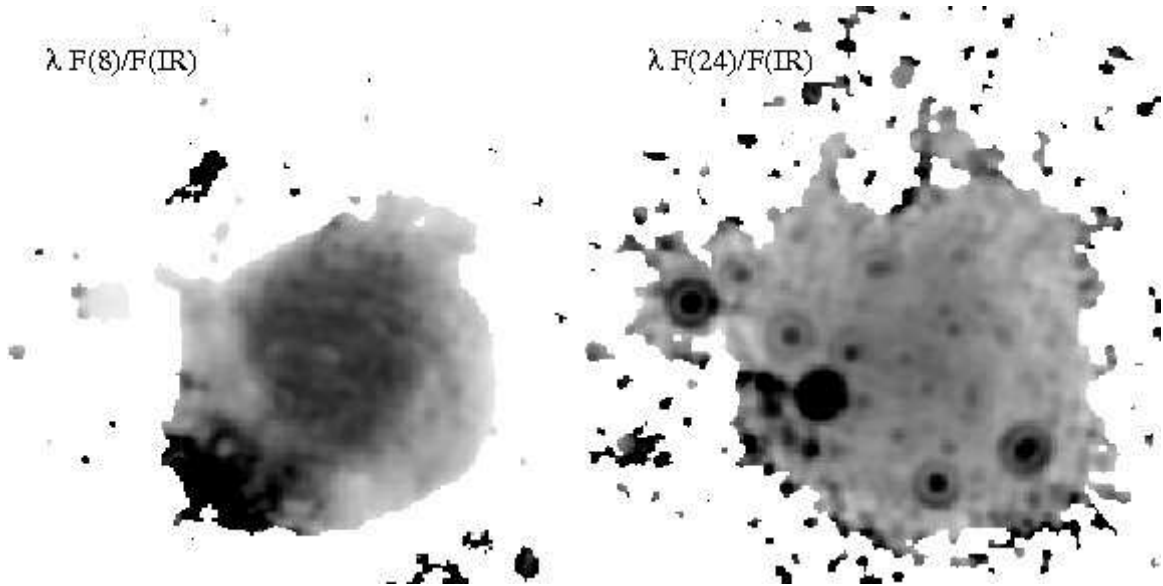}
\caption{The images of M101 in $\lambda F(8~\micron)/F(IR)$ and
$\lambda F(24~\micron)/F(IR)$ are shown.  The images are displayed in
reverse B/W with a linear stretch from 0--0.30 and 0--15,
respectively.  This same region displayed in
Fig.~\ref{fig_m101_truecol} is shown (30$\arcmin$ roughly centered on
M101). \label{fig_m101_160res} }
\end{figure*}

Finally, the aromatic feature and underlying continuum emission can be
compared to that of the large grains by including the longer
wavelength MIPS images in the analysis.  Fig.~\ref{fig_m101_160res}
shows two images: the first is the ratio of the 8~\micron\ flux to the
total infrared ($\lambda F(8~\micron)/F(IR)$) and the second is the
ratio of the 24~\micron\ flux to the total infrared ($\lambda
F(24~\micron)/F(IR)$).  Generally, the 8~\micron\ emission traces the
aromatic features (with a minor contribution from the continuum), the
24~\micron\ emission traces the hot dust emission from small grains,
and the total infrared traces the bulk of the dust grains which
dominate the dust mass and have a peak emission around 160~\micron\
(see Fig.~\ref{fig_global_sed}).  To first order, $\lambda
F(8~\micron)/F(IR)$ and $\lambda F(24~\micron)/F(IR)$ trace the ratio
of aromatic grain mass to dust mass and small grain mass to dust mass,
respectively, modified by the illuminating radiation field.  All the
images used were first convolved to the 160~\micron\ resolution of
40\arcsec\ (1.3~kpc at 6.7~Mpc) using the kernels described in
\S\ref{sec_res_match}.  The total infrared flux was calculated by a
simple integration of the IRAC 8~\micron\ through MIPS 160~\micron\
images.  The $\lambda F(8~\micron)/F(IR)$ image is dominated by a high
ratio in the inner region with a fairly abrupt drop to much lower
values.  The brightest HII regions are seen to have shallow
depressions compared to other regions at the same radii.  The $\lambda
F(24~\micron)/F(IR)$ image displays a different morphology with a
relatively flat ratio overall with strong enhancements in the HII
regions.  The depression of the aromatic emission is not due to lack
of exciting photons as these are clearly present given the enhancement
of the 24~\micron\ emission.  The depression of the aromatic emission
in HII regions must then be due to processing of the carriers of these
features.  The behavior of these two ratios supports the results of
the comparison to the H$\alpha$ given in the previous paragraph.  The
variations of the 8~\micron\ emission are dominated by variations in
ionization while the 24~\micron\ emissions are dominated by the
radiation field intensity (as traced by the H$\alpha$ emission).

\section{Discussion}

\begin{figure*}[tbp]
\epsscale{1.1}
\plotone{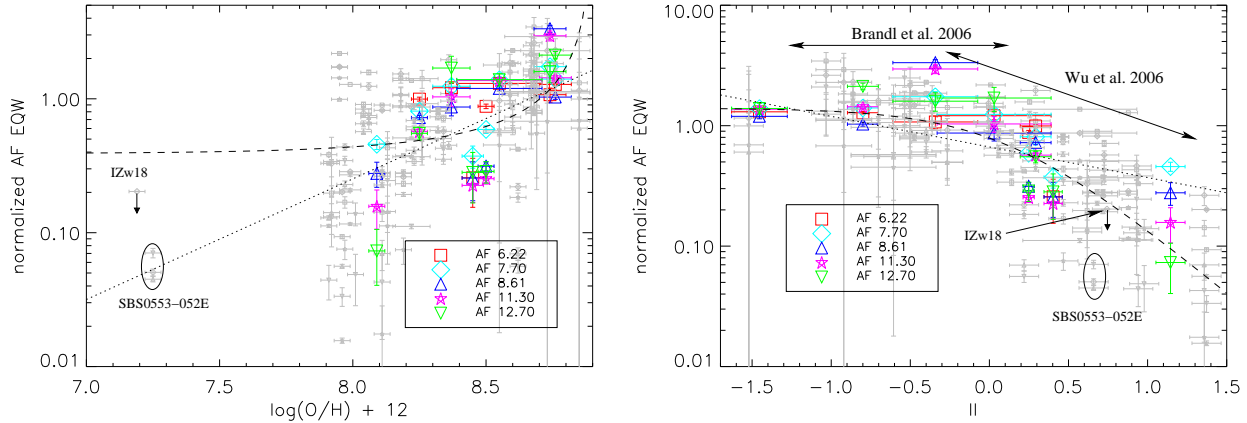}
\caption{The normalized equivalent widths of the 5 highest equivalent
width aromatic features (AFs) are plotted versus metallicity (left)
and ionization index (II, right).  These two plots are the
same as shown in Fig.~\ref{fig_spec_eqw} but now include the starburst
sample from \citet{Engelbracht08SB} in grey.  The locations of the two
most metal poor galaxies (SBS0335-052E and IZw18) are labeled.
In addition, the II
ranges probed in \citet{Brandl06SB} and \citet{Wu06} are indicated.  The
normalization was done the same as in Fig.~\ref{fig_spec_eqw}.  
\label{fig_comp_lit} }
\end{figure*}

The behavior of the aromatic feature equivalent widths versus
metallicity and II seen in this paper for M101 HII regions is also
seen for starburst galaxies.  This is shown in
Fig.~\ref{fig_comp_lit} which combines the measurements for the M101
HII regions presented in this paper with the starburst galaxy
measurements from \citet{Engelbracht08SB}.  Both samples show a better
correlation of the aromatic equivalent widths versus II than
metallicity.  The combined M101 HII region and starburst galaxy
samples probe ionization indexes between -1.5 and 1.4  
(equivalent to [NeIII]/[NeII] ratios of 0.03--25) and metallicities
(log(O/H)+12) from 7.9--8.8.  There are two galaxies (SBS0335-052E and
IZw18) which have metallicities below 7.9, but this is not
enough points to statistically probe the behavior in this range.  It is interesting
to note that these two galaxies do not have the most extreme II values
and fit fairly well in the trend of II versus aromatic feature
equivalent widths.  The residuals of fit of II versus aromatic feature
equivalent width were plotted versus metallicity to see if the scatter
around the fit was due to metallicity, but no trend was
present.

The correlation of the aromatic equivalent widths with II is not a
simple power law, but shows a change at around a II value of zero.
Below this value, the equivalent widths are fairly constant indicating
that the material responsible for the aromatic features is as robust
as the material responsible for the underlying dust continuum
emission.  Above this value, the equivalent widths drop significantly
with increasing II.  This implies that the material responsible for
the aromatic features is being modified/destroyed at a faster rate
than the material responsible for the underlying dust continuum
emission.

The complex behavior of the aromatic
equivalent widths with II explains the seemingly contradictory
results found by previous authors.  \citet{Brandl06SB} studied low
ionizations ($-1.25 < \mbox{II} < 0.05$) and found no
variation.  \citet{Wu06} studied high ionizations
($-0.3 < \mbox{II} < 1.3$) and found large variations.  Only by
studying the full range of II values does the complex nature of the
correlation between II and aromatic equivalent width emerge.

The complex behavior of the aromatic equivalent widths also explains
why the bright point sources (e.g., HII regions) in M101 change fairly
abruptly from yellow to red at approximately the same radii in
Fig.~\ref{fig_m101_truecol}.  The colors of the HII regions are
primarily composed of green (IRAC 8~\micron) and red (MIPS
24~\micron).  In the inner regions of M101, the HII regions have high
metallicity and low ionization (e.g., Fig.~\ref{fig_spec_diag}c).
This means that they have II values below 0 and, thus, normal aromatic
equivalent widths (e.g., IRAC 8~\micron\ emission).  In the outer
regions, the HII regions have low metallicity and high
ionization. Thus, they have much weaker aromatic emission which
strongly reduces the green color.  Red color in this particular type
of 3 color image is a strong indication of weak aromatic emission.

Our finding that the aromatic equivalent widths in the M101 HII
regions and starburst galaxies correlates better with a measure of
processing (ionization) 
than with a measure of formation (metallicity) is not too surprising.
There has been ample evidence for such processing in spatially
resolved studies in reflection nebulae and HII regions in the Milky
Way \citep{Cesarsky00, Berne07}.
But until this work, it is not been clear if the spatially resolved
processing of the aromatic carriers also applies to the global spectra
of massive star forming regions (i.e., HII regions and starburst
galaxies).   This work indicates that the processing of the aromatic
carriers is happening on a global scale in massive star forming
regions.

\begin{figure}[tbp]
\epsscale{1.2}
\plotone{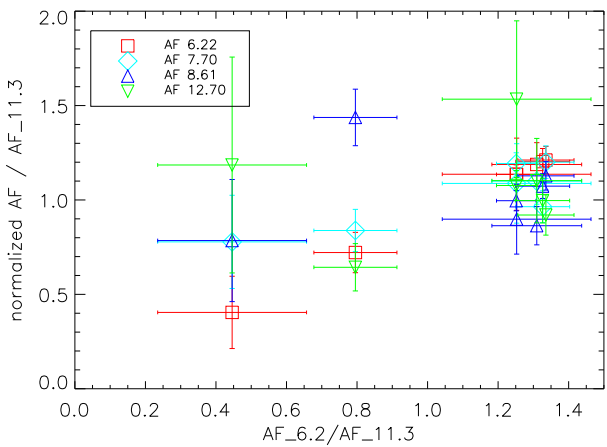}
\caption{The ratio of the strength of the 4 strongest
aromatic features (AFs) to the 11.3 AF are plotted versus the 6.2/11.3
AFs.  The normalization was done to
the average ratio of each AF.  The normalization values are 1.01,
3.41, 0.80, and 0.57 for the 6.22, 7.70, 8.61, 
and 12.70~\micron\ aromatic features, respectively.  The points for
the 6.22 AF defined the one-to-one correspondence.
\label{fig_spec_ratio_x611} }
\end{figure}

Models where PAH molecules are the carriers of the aromatic features
would predict a correlation of the ratios of different
aromatic features with ionization index if the processing
mechanisms were ionization or size selective destruction.  The lack of
such a correlation (see Fig.~\ref{fig_spec_ratio}) seems to imply that
these two processing mechanisms may not be the dominant cause of the
strong correlation seen with ionization index.
Fig.~\ref{fig_spec_ratio_x611} plots the ratio of the strongest 4
aromatic features and the 11.3~\micron\ aromatic versus the 6.2/11.3 aromatic
ratio.  In PAH models of the aromatic features \citep{Galliano06}, the
6.2~\micron\ aromatic traces small, ionized PAHs and the 11.3~\micron\ aromatic traces
larger, neutral 
PAHs.  Thus, the 6.2/11.3 ratio traces ionized/neutral and/or
small/large PAHs.  The 7.70~\micron\ and 8.61~\micron\ aromatics also trace small,
ionized PAHs and the 12.7~\micron\ larger, neutral PAHs.  While there is
scatter in Fig.~\ref{fig_spec_ratio_x611}, the 7.70/11.3 and 8.61/11.3
ratios correlate fairly well with the 6.2/11.3 ratio.  On the other
hand, the 12.7/11.3 does not correlate at all with the 6.2/11.2
ratio.  Thus, the behavior of the relative strengths of the 5
strongest aromatic features is consistent with the PAH molecule
model.  The lack of correlation seen between the variations in the
aromatic features with ionization
(Fig.~\ref{fig_spec_ratio}) is not consistent with the PAH molecule
model.  

There seem to be two different effects at work in HII regions.  The
first is consistent with the PAH molecule model and explains the
behavior of the spatially resolved aromatic feature ratios.  The
second is required to 
explain the good correlation between the decrease in the globally
integrated aromatic
equivalent widths and increasing ionization {\em and} the 
lack of a correlation between the aromatic feature ratios and
ionization index.  This second effect may be related to the
small grain model for the aromatic features.  If small grains can
produce aromatic features and provide the parent population of the PAH
molecules in photodissociation regions regions (PDRs) \citep{Berne07}
then it may be that the 
decrease in the strength of the aromatic equivalent width is due to
processing of the small grains.  Another possible explanation could be
that the sizes of the PDRs in regions with harder radiation fields
(and, thus, higher ionizations) are significantly reduced.  As the
aromatic features are known to emerge predominately from the PDR
regions surrounding the HII regions and the underlying continuum
emission emerges from the whole HII region, decreasing the area of the
PDR would result in smaller aromatic equivalent widths.  Decreasing
PDR areas with lower metallicities may also explain the low CO emission
seen in low metallicity starburst galaxies \citep{Leroy07}.

The processing of the aromatic feature carriers by massive star
formation is one line of evidence that such star formation actively
modifies the properties of the surrounding dust grains.
There is additional evidence that active, high-mass star
formation modifies the properties of nearby dust grains.  One
observational signature of this effect is the lack of the 2175~\AA\
extinction bump in starburst galaxies \citep{Calzetti94, Gordon97}.
Ultraviolet (UV) extinction curves studies in the Small Magellanic
Cloud \citep{Gordon98}, Large Magellanic Cloud \citep{Misselt99}, and
Milky Way \citep{Clayton00, Valencic03} paint a consistent picture
where the 2175~\AA\ bump is depressed and the overall UV extinction
increased in regions near massive star formation \citep{Gordon03}.
These changes in the UV extinction can be explained by the removal of
the small graphite grains that produce the 2175~\AA\ feature, along
with an
increase in the number of small silicate grains that are responsible
for the majority of the ultraviolet extinction \citep{Weingartner01,
Clayton03}.  Given that these changes are in the small grain
populations of the dust, it is not surprising that similar changes are
seen in the mid-infrared where the small grains
emit their absorbed energy.  It is not clear if the changes seen
in the aromatic features are directly correlated with the changes seen
in the 2175~\AA\ feature.  More work in both the ultraviolet and
mid-infrared is needed to answer this question.

\section{Summary and Conclusions}

We have studied trends in aromatic feature
behavior with metallicity (log(O/H)+12 = 7.9--8.8) and 
ionization (II = -1.5--1.4 or [NeIII]/[NeII] = 0.03--25). Over this
range, we have 
found that the equivalent width of the aromatic features is
better correlated with the ionization index (II) than metallicity
(log(O/H)+12) in both the M101 regions studied in this paper as well
as the starburst galaxies presented in \citet{Engelbracht08SB}.  The
data available for 
metallicities between log(O/H)+12 = 7.1--7.9 are not sufficient to
probe these trends for such low metallicities.
Our result implies that the decrease in aromatic equivalent widths 
seen in regions of massive star formation (e.g., HII regions and
starburst galaxies) is primarily due to processing, not formation.
This is not to say that formation does 
not play a role, but just that it is not the dominant effect.  The finding
of processing of the aromatic carriers is consistent with existing
evidence that the 2175~\AA\ extinction feature carrier is processed in
regions of massive star formation.

We found that the correlation of the aromatic equivalent widths with
II is not a single power law as has been proposed in
previous works.  Instead the aromatic equivalent widths are constant
till a threshold II of $\sim$0 and then decrease.  This decrease is
well fit with a power law.  This more complex behavior is consistent
with previous work 
\citep{Brandl06SB} and \citep{Wu06} which probed a
narrower range of II and seemed to find conflicting results.  Only
the large range in both II and metallicity probed in this work
and the companion work of \citet{Engelbracht08SB} allowed for the true
nature of the aromatic feature variations to be revealed.

The lack of a clear dependence of the ratio of different aromatic
features to the 11.3 aromatic feature on II
calls into question either ionization or size selective destruction of
PAH molecules as the mechanism for the observed reduction in the
aromatic equivalent width versus ionization.  This might
indicate a 
more solid material origin for the aromatics \citep{Berne07} which
dominate the global measurements of HII regions.  Or it might imply
that the area of the PDR regions decreases with respect to the area of the small
grain emission as the radiation field hardness (and thus
ionization) increases.

\acknowledgements
This work is based on observations made
with the {\em Spitzer Space Telescope}, which is operated by the Jet
Propulsion Laboratory, California Institute of Technology under NASA
contract 1407. Support for this work was provided by NASA through
Contract Number \#1255094 issued by JPL/Caltech to the University of
Arizona.

\clearpage
\begin{landscape}

\begin{deluxetable*}{lcccccccc}
\tablewidth{0pt}
\tablecaption{Atomic Emission Line Strengths\tablenotemark{a} \label{tab_atomic_strength}}
\tablehead{ \colhead{name}  & \colhead{[ArII]} & \colhead{[ArIII]} & \colhead{[SIV]} & \colhead{[NeII]} & \colhead{[NeIII]} & \colhead{[SIII] 18} & \colhead{[SIII] 33} & \colhead{[SiII]} \\
 & \colhead{7.0~\micron} & \colhead{9.0~\micron} & \colhead{10.5~\micron} & \colhead{12.8~\micron} & \colhead{15.5~\micron} & \colhead{18.7~\micron} & \colhead{33.5~\micron} & \colhead{34.8~\micron} }
\startdata
Nucleus  & $    3.30 \pm     0.51$  & \nodata & \nodata & $    7.57 \pm     0.51$  & $    1.20 \pm     0.25$  & $    4.50 \pm     0.51$  & $   18.78 \pm     0.89$  & $   20.26 \pm     0.79$  \\
Hodge~602 & $    1.00 \pm     0.23$  & \nodata & $    0.34 \pm     0.12$  & $    2.30 \pm     0.16$  & \nodata & $    1.54 \pm     0.13$  & $    3.74 \pm     0.12$  & $    3.48 \pm     0.14$  \\
Searle~5  & $    1.04 \pm     0.17$  & $    0.09 \pm     0.09$  & $    0.18 \pm     0.05$  & $    4.13 \pm     0.16$  & $    0.16 \pm     0.11$  & $    4.01 \pm     0.18$  & $   20.36 \pm     0.28$  & $    9.14 \pm     0.18$  \\
NGC~5461  & $    1.78 \pm     5.33$  & $   10.74 \pm     0.81$  & $   28.39 \pm     0.81$  & $   27.32 \pm     1.54$  & $   56.09 \pm     5.82$  & $   60.56 \pm     3.94$  & $  259.21 \pm     5.53$  & $   80.58 \pm     3.67$  \\
NGC~5462  & $    0.86 \pm     0.53$  & $    2.25 \pm     0.50$  & $    2.75 \pm     1.21$  & $    4.34 \pm     7.83$  & $    6.50 \pm     1.95$  & $    7.46 \pm     1.49$  & $   43.90 \pm     1.02$  & $   15.01 \pm     0.70$  \\
NGC~5455  & $    0.34 \pm     1.61$  & $    1.25 \pm     0.09$  & $    3.08 \pm     0.09$  & $    3.85 \pm     0.15$  & $   11.19 \pm     3.19$  & $    7.93 \pm     1.68$  & $   46.14 \pm     0.74$  & $   18.63 \pm     0.59$  \\
NGC~5447  & $    0.25 \pm     0.11$  & $    0.92 \pm     0.11$  & $    3.33 \pm     0.10$  & $    1.61 \pm     0.20$  & $    4.27 \pm     0.56$  & $    5.41 \pm     0.64$  & $   27.51 \pm     1.31$  & $   10.60 \pm     1.07$  \\
NGC~5471  & \nodata & $    0.52 \pm     0.15$  & $   12.92 \pm     0.73$  & $    0.96 \pm     0.29$  & $   20.63 \pm     2.33$  & $    8.47 \pm     1.38$  & $   32.60 \pm     0.75$  & $   11.27 \pm     0.57$  \\
\enddata
\tablenotetext{a}{Units are 10$^{-17}$ W m$^{-2}$.}
\end{deluxetable*}

\begin{deluxetable*}{lccccccccccc}
\tablewidth{0pt}
\tablecaption{Aromatic Emission Line Strengths\tablenotemark{a} \label{tab_aromatic_strength}}
\tablehead{ \colhead{name}  & \colhead{5.7~\micron} & \colhead{6.2~\micron} & \colhead{7.7~\micron} & \colhead{8.3~\micron} & \colhead{8.6~\micron} & \colhead{10.7~\micron} & \colhead{11.3~\micron} & \colhead{12.0~\micron} & \colhead{12.7~\micron} & \colhead{14.0~\micron} & \colhead{17.0~\micron} }
\startdata
Nucleus  & $    7.86 \pm     1.87$  & $   77.06 \pm     1.78$  & $  271.43 \pm     7.87$  & $   23.36 \pm     2.00$  & $   56.62 \pm     2.14$  & $    1.64 \pm     0.47$  & $   61.58 \pm     2.36$  & $   15.98 \pm     1.07$  & $   36.27 \pm     2.27$  & $    1.04 \pm     0.57$  & $   55.12 \pm     3.72$  \\
Hodge~602 & $    2.53 \pm     1.25$  & $    9.61 \pm     1.16$  & $   37.42 \pm     3.83$  & $    1.76 \pm     0.90$  & $   16.03 \pm     0.95$  & $    1.37 \pm     0.24$  & $   12.08 \pm     1.03$  & $    1.63 \pm     0.38$  & $    4.25 \pm     0.74$  & \nodata & $    4.97 \pm     0.42$  \\
Searle~5  & \nodata & $   12.54 \pm     0.44$  & $   41.64 \pm     2.20$  & $    1.03 \pm     0.49$  & $    9.77 \pm     0.45$  & $    0.02 \pm     0.13$  & $    9.39 \pm     0.45$  & $    1.88 \pm     0.22$  & $    4.73 \pm     0.50$  & $    0.59 \pm     0.14$  & $    1.99 \pm     0.30$  \\
NGC~5461  & $    6.64 \pm     1.25$  & $   66.21 \pm     2.58$  & $  205.47 \pm    13.74$  & $   12.74 \pm     3.50$  & $   40.30 \pm     3.08$  & $    2.48 \pm     1.01$  & $   50.56 \pm     4.50$  & $   14.73 \pm     2.41$  & $   30.49 \pm     5.56$  & \nodata & $    3.21 \pm     4.16$  \\
NGC~5462  & $    1.46 \pm     1.14$  & $   20.21 \pm     1.54$  & $   64.84 \pm     7.92$  & $    4.42 \pm     1.86$  & $   13.38 \pm     1.89$  & $    1.34 \pm     0.64$  & $   16.13 \pm     2.42$  & $    1.88 \pm     1.30$  & $   13.54 \pm     3.05$  & $    2.95 \pm     0.71$  & $    0.72 \pm     1.18$  \\
NGC~5455  & $    0.78 \pm     0.64$  & $   14.42 \pm     0.48$  & $   38.75 \pm     2.86$  & $    2.84 \pm     0.49$  & $   10.77 \pm     0.42$  & $    0.39 \pm     0.16$  & $   10.87 \pm     0.51$  & $    2.34 \pm     0.32$  & $    5.93 \pm     0.65$  & $    0.63 \pm     0.21$  & \nodata \\
NGC~5447  & $    0.50 \pm     0.44$  & $    1.30 \pm     0.52$  & $    8.41 \pm     1.58$  & $    1.35 \pm     0.74$  & $    2.12 \pm     0.68$  & \nodata & $    2.93 \pm     0.75$  & $    0.54 \pm     0.37$  & $    1.90 \pm     0.78$  & \nodata & $    2.67 \pm     1.24$  \\
NGC~5471  & $    0.24 \pm     0.69$  & \nodata & $    8.82 \pm     0.76$  & \nodata & $    2.15 \pm     0.46$  & $    1.27 \pm     0.29$  & $    2.85 \pm     0.92$  & \nodata & $    0.84 \pm     0.38$  & $    0.74 \pm     0.49$  & $    0.58 \pm     1.73$  \\
\enddata
\tablenotetext{a}{Units are 10$^{-17}$ W m$^{-2}$.}
\end{deluxetable*}

\begin{deluxetable*}{lccccccccccc}
\tablewidth{0pt}
\tablecaption{Aromatic Emission Line Equivalent Widths\tablenotemark{a} \label{tab_aromatic_eqw}}
\tablehead{ \colhead{name}  & \colhead{5.7~\micron} & \colhead{6.2~\micron} & \colhead{7.7~\micron} & \colhead{8.3~\micron} & \colhead{8.6~\micron} & \colhead{10.7~\micron} & \colhead{11.3~\micron} & \colhead{12.0~\micron} & \colhead{12.7~\micron} & \colhead{14.0~\micron} & \colhead{17.0~\micron} }
\startdata
Nucleus  & $    0.16 \pm     0.04$  & $    1.44 \pm     0.03$  & $    4.91 \pm     0.14$  & $    0.46 \pm     0.04$  & $    1.17 \pm     0.04$  & $    0.05 \pm     0.01$  & $    1.90 \pm     0.07$  & $    0.51 \pm     0.03$  & $    1.18 \pm     0.07$  & $    0.03 \pm     0.02$  & $    1.46 \pm     0.10$  \\
Hodge~602 & $    0.29 \pm     0.14$  & $    1.21 \pm     0.15$  & $    6.09 \pm     0.62$  & $    0.35 \pm     0.18$  & $    3.80 \pm     0.23$  & $    0.54 \pm     0.09$  & $    3.92 \pm     0.34$  & $    0.41 \pm     0.10$  & $    0.89 \pm     0.16$  & \nodata & $    0.59 \pm     0.05$  \\
Searle~5  & \nodata & $    1.47 \pm     0.05$  & $    4.85 \pm     0.26$  & $    0.13 \pm     0.06$  & $    1.36 \pm     0.06$  & $    0.00 \pm     0.03$  & $    1.81 \pm     0.09$  & $    0.33 \pm     0.04$  & $    0.78 \pm     0.08$  & $    0.08 \pm     0.02$  & $    0.17 \pm     0.03$  \\
NGC~5461  & $    0.13 \pm     0.02$  & $    0.99 \pm     0.04$  & $    2.07 \pm     0.14$  & $    0.12 \pm     0.03$  & $    0.36 \pm     0.03$  & $    0.02 \pm     0.01$  & $    0.34 \pm     0.03$  & $    0.09 \pm     0.01$  & $    0.16 \pm     0.03$  & \nodata & $    0.01 \pm     0.01$  \\
NGC~5462  & $    0.11 \pm     0.08$  & $    1.36 \pm     0.10$  & $    4.35 \pm     0.53$  & $    0.32 \pm     0.13$  & $    0.99 \pm     0.14$  & $    0.12 \pm     0.06$  & $    1.37 \pm     0.21$  & $    0.15 \pm     0.10$  & $    0.95 \pm     0.21$  & $    0.15 \pm     0.04$  & $    0.02 \pm     0.03$  \\
NGC~5455  & $    0.07 \pm     0.06$  & $    1.12 \pm     0.04$  & $    2.83 \pm     0.21$  & $    0.21 \pm     0.04$  & $    0.82 \pm     0.03$  & $    0.03 \pm     0.01$  & $    0.75 \pm     0.04$  & $    0.14 \pm     0.02$  & $    0.31 \pm     0.03$  & $    0.02 \pm     0.01$  & \nodata \\
NGC~5447  & $    0.14 \pm     0.12$  & $    0.29 \pm     0.12$  & $    1.31 \pm     0.25$  & $    0.19 \pm     0.11$  & $    0.29 \pm     0.09$  & \nodata & $    0.30 \pm     0.08$  & $    0.05 \pm     0.03$  & $    0.16 \pm     0.06$  & \nodata & $    0.10 \pm     0.04$  \\
NGC~5471  & $    0.08 \pm     0.22$  & \nodata & $    1.60 \pm     0.14$  & \nodata & $    0.32 \pm     0.07$  & $    0.11 \pm     0.03$  & $    0.21 \pm     0.07$  & \nodata & $    0.04 \pm     0.02$  & $    0.03 \pm     0.02$  & $    0.01 \pm     0.03$  \\
\enddata
\tablenotetext{a}{Units \micron.}
\end{deluxetable*}

\clearpage
\end{landscape}

\end{document}